\documentclass[%
 reprint,
 amsmath,amssymb,
 aps,
]{revtex4-2}

\usepackage{graphicx}
\usepackage{dcolumn}
\usepackage{bm}
\usepackage{hyperref}
\usepackage[braket, qm]{qcircuit}
\usepackage{tikz}
\usepackage{pgfplots}
\usepackage{placeins}
\usepackage{tikz-3dplot}
\usepackage{xcolor}
\usepackage{float}
\usepackage{subfigure}
\usepackage[normalem]{ulem}

\begin{document}


\title{Quantum simulation of a qubit with non-Hermitian Hamiltonian}
\author{Anastashia Jebraeilli}
 \email{anaj@uga.edu}
\author{Michael R. Geller}
\affiliation{
 Department of Physics and Astronomy, University of Georgia, \\
 Center for Simulational Physics, University of Georgia, \\
Athens, Georgia 30602, USA
}

\date{February 19, 2025}

\begin{abstract}
Modeling non-Hermitian Hamiltonians is increasingly important in classical and quantum domains, especially when studying open systems, $PT$ symmetry, and resonances. However, the quantum simulation of these models has been limited by the extensive resources necessary in iterative methods with exponentially small 
postselection success probability. Here we employ a fixed-depth variational circuit to circumvent these limitations, enabling simulation deep into the $PT$-broken regime surrounding an exceptional point. Quantum simulations are carried out using IBM superconducting qubits. The results underscore the potential for variational quantum circuits and machine learning to push the boundaries of quantum simulation, offering new methods for exploring quantum phenomena with near-term intermediate-scale quantum technology.
\end{abstract}

\maketitle

\section{\label{sec:introduction}Introduction}

Open quantum systems are not restricted to Hermitian Hamiltonian dynamics. A diverse array of systems—including superconducting qubits, optical waveguides, oscillators, polarized photons, ultracold Fermi gases, and atomic or molecular configurations—exhibit behavior that can be described by non-Hermitian Hamiltonians \cite{ElGanainy2018, Rueter2010, Biesenthal2019, Klauck2019, Xiao2017, Zheng2013, Wen2019, Quijandria2018, Naghiloo_2019, Bender2013, Pick2019, Wu2019, Regensburger2012}. Despite their prevalence, non-Hermitian Hamiltonians have yet to be simulated on near-term intermediate-scale quantum (NISQ) devices. The simulation of non-Hermitian Hamiltonians is not only interesting from a quantum computing perspective but also can help 
unlock a deeper understanding of quantum mechanics itself \cite{bender1998pt, Bender_2007, Bender_2003, bender2023ptsymmetric, PhysRevLett.89.270401, Bender_1999, Bender_2005}.

\begin{figure}
\begin{center}
\includegraphics[width=8cm]{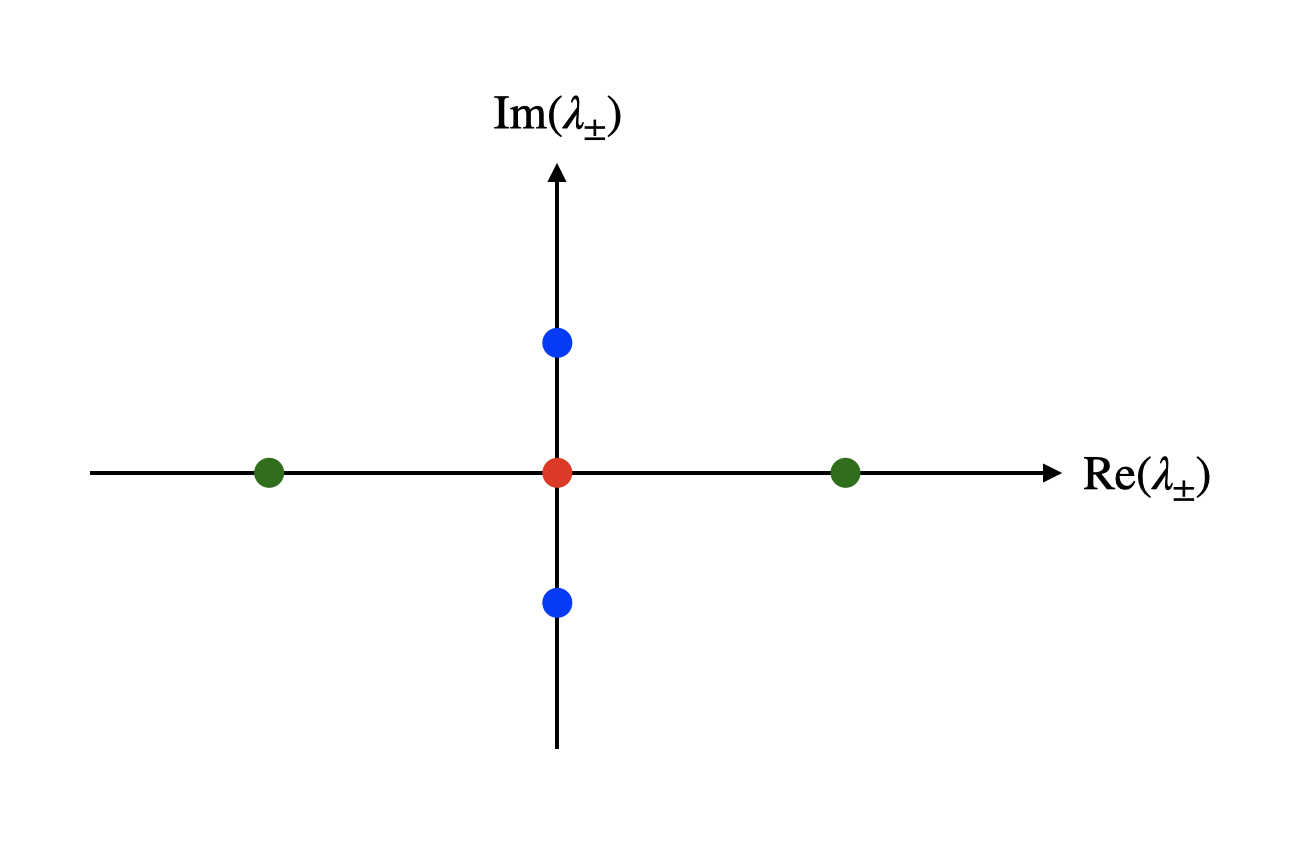} 
\caption{Eigenvalues of $H_{\rm eff}$ for 
$| \omega | > \Gamma$ (green points) and 
$| \omega | < \Gamma$ (blue points).  At the exceptional points $\omega = \pm\Gamma $,  the eigenvalues vanish
(red point). } 
\label{spectrum figure}
\end{center}
\end{figure} 

Non-Hermicity results in the inclusion of phenomena that are absent in closed Hermitian systems. Exceptional points (EPs), where the eigenvectors of a system coalesce, are examples of such non-Hermitian behavior, along with $PT$ symmetry breaking. 

Recently, a method using postselection has been proposed \cite{Zhang2021} to simulate the non-Hermitian Hamiltonian  
\begin{equation}
H_{\text{eff}} = 
 \frac{\omega}{2}\sigma_x + \frac{i \Gamma}{2} \sigma_z 
 =  \begin{pmatrix}  \frac{i \Gamma}{2}  &  \frac{\omega}{2}  \\
\frac{\omega}{2}  & - \frac{i \Gamma}{2}
 \end{pmatrix} , 
 \ \ \ {\rm EPs} \! : \omega = \pm  \Gamma,
 \label{Heff def}
\end{equation}
where $\sigma_x$ and $\sigma_z$ are Pauli matrices. Here $\Gamma \ge 0$, while $\omega \in {\mathbb R}$ is arbitrary.  
The eigenvalues and eigenvectors of $ H_{\text{eff}} $ respectively are
\begin{align}
\lambda_{\pm} &= \pm \frac{\sqrt{\omega^2 - \Gamma^2}}{2} , 
\label{equation2a} \\
|\phi_\pm\rangle &= c_\pm
\begin{pmatrix} \omega \\ -i\Gamma \pm \sqrt{\omega^2 - \Gamma^2} \end{pmatrix} \! , 
\label{equation2b}
\end{align}
where
$ c_\pm = (\omega^2 + |\! -i\Gamma \pm \sqrt{\omega^2 - \Gamma^2} \, |^2 )^{-\frac{1}{2}}$ is real and positive.
In the special (and trivial) case where $\Gamma = 0$, standard Hamiltonian simulation is recovered; henceforth, we will assume $\Gamma > 0.$ If $|\omega| > \Gamma$, both eigenvalues are real, and if $|\omega| < \Gamma$ the eigenvalues become a purely imaginary complex conjugate pair, as illustrated in Fig.~\ref{spectrum figure}.

An exceptional point (EP) in parameter space occurs if the Hamiltonian becomes defective (lacking two linearly independent eigenvectors).
The model (\ref{Heff def}) has two EPs, at
$\omega = \pm \Gamma$. At an EP, the single eigenvector is
\begin{eqnarray}
 | \Phi_{\omega=\Gamma} \rangle  = \frac{1}{\sqrt{2}} 
\begin{pmatrix} 1 \\ -i \end{pmatrix} \ \ {\rm or} \ \ 
 | \Phi_{\omega=-\Gamma} \rangle  = \frac{1}{\sqrt{2}} 
\begin{pmatrix} 1 \\ i \end{pmatrix}, 
\label{ep eigenvector}
\end{eqnarray}
a pure state on the equator of the Bloch sphere.

We see that, in contrast to a Hermitian Hamiltonian,
 $H_{\rm eff}$ can have complex eigenvalues. A standard framework for understanding this is through $PT$ symmetry, where $P$ is parity 
 (satisfying $P {\vec r} P^{-1} = -{\vec r} $)
 and $T$ is time reversal (satisfying $TzT^{-1} = z^{*}$ where $z \in {\mathbb C}$).  A non-Hermitian Hamiltonian that commutes with $PT$ is defined as $PT$-invariant. A $PT$-invariant Hamiltonian with a real spectrum is classified as 
 $PT$-symmetric, whereas $PT$ symmetry is broken if the eigenvalues have imaginary parts.

To make use of $PT$ symmetry in the qubit model (\ref{Heff def}), we need to define a two-dimensional unitary parity operator $P$. A natural choice is $P = \sigma^x$, which satisfies $P \sigma^z P^{-1} = - \sigma^z$ and renders the model $PT$-invariant \cite{Dogra2021}.
If $ \omega < - \Gamma $ or $ \omega > \Gamma $, the model is in a $PT$-symmetric regime; if  
$ - \Gamma < \omega <  \Gamma $ the $PT$ symmetry is broken. The EPs at $ \omega  = \pm \Gamma $ separate these regimes.

Suppose a Hamiltonian $H$ is $PT$-invariant, and let $| \phi \rangle$ be an eigenfunction with eigenvalue $\lambda$.
 Then, from $ [H , PT ] = 0$, we see that 
 $| \phi^\prime \rangle :=  PT | \phi \rangle$ is also an $H$ eigenfunction with eigenvalue $\lambda$, potentially leading to twofold degeneracy.
If  $ | \phi^\prime \rangle  =  e^{i \gamma} | \phi \rangle $ with $\gamma \in {\mathbb R}$, then 
$ | \phi^\prime \rangle $ is not distinct from  $| \phi \rangle $ and there is no energy degeneracy resulting from the $PT$-invariance, but $| \phi \rangle $ is simultaneously an eigenfunction of $PT$ with eigenvalue $e^{i \gamma}$:
\begin{eqnarray}
H  | \phi \rangle  =  \lambda | \phi \rangle, \ \ 
PT | \phi \rangle  =  e^{i \gamma} | \phi \rangle .
\end{eqnarray}
However, if there is no $\gamma \in {\mathbb R}$ such that $ | \phi^\prime \rangle  =  e^{i \gamma} | \phi \rangle $, then
$  | \phi \rangle$ and $ PT | \phi \rangle$ are distinct, and the $PT$ invariance leads to twofold energy degeneracy.
This has an important consequence for the $PT$-invariant model (\ref{Heff def}) at an EP. Namely, the eigenvectors (\ref{ep eigenvector}) are $PT$ eigenfunctions:
\begin{eqnarray}
PT | \Phi_{\omega=\pm \Gamma} \rangle = | \Phi_{\omega=\pm\Gamma} \rangle  .
\end{eqnarray}

A circuit that simulates $H_{\rm eff}$ is given in Fig.~\ref{fig:model}, as we explain below.
By entangling system qubits with ancillary qubits (one per system qubit), the Zhang \textit{et al.} method simulates the entanglement of a quantum system with its surroundings leading to (\ref{Heff def}). 

\begin{figure}
\begin{center}
\mbox{
\Qcircuit @C=1em @R=.7em {
 \lstick{\ket{\rm in}} & \gate{R_x(\theta)}    &  \ctrl{1}  & \qw & \qw  & \qw    & & & \lstick{\ket{\rm out}}    \\
\lstick{\ket{0}}  & \qw & \gate{R_x(\phi)}  & \qw  &  \measureD{\mbox{$x_2$}} & & x_2=0     \\
}}
\end{center}
  \caption{\label{fig:model} Single step of the quantum circuit for simulating \(H_{\text{eff}}\), proposed by Zhang \textit{et al.} \cite{Zhang2021}. Here $R_x(\varphi) = e^{-\frac{i}{2} \varphi \sigma^x}$ is an $x$ rotation and $|\phi| \ll 1$.}
  \label{fig:model}
\end{figure}
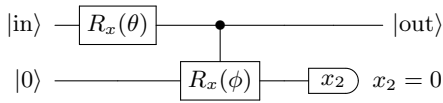

\subsection{\label{sec:nonhermsim}Simulating non-Hermicity}

We assume that the system qubit (qubit 1, the upper qubit in Fig.~\ref{fig:model}) is initially in a pure state $\ket{\psi} = \psi_0 \ket{0} + \psi_1\ket{1}$ with 
$\| \ket{\psi} \| = 1$.
We want to implement the map
\begin{eqnarray}
\ket{\psi} \mapsto \frac{ A_t \ket{\psi} }{\| A_t \ket{\psi} \! \| }, \ \ 
A_t := e^{-i H_{\rm eff} t} , \ \ t \ge 0,
\end{eqnarray}
where $A_t$ is a linear (but not necessarily unitary) operator. To simulate (\ref{Heff def}), let $\tau = t/K$, where $K$ is a positive integer (number of Trotter steps), and note that 
\begin{eqnarray}
A_t = e^{-i (-i \Gamma \, e_{11} + \frac{\omega}{2} \sigma^x)t} = \lim_{K \rightarrow \infty} A_{\tau}^{K},
\end{eqnarray}
where $e_{11} = |1\rangle \langle 1 |$. For large $K$,
\begin{eqnarray}
A_{\tau} = e^{-\Gamma \tau \, e_{11}} 
R_x(\omega \tau) .
\label{trotter step}
\end{eqnarray}

In the approach of Zhang \textit{et al.} \cite{Zhang2021}, the Trotter step $A_{\tau}$ is implemented by the circuit shown in Fig. \ref{fig:model} with $|\phi|\ll  1.$ Each step of the circuit applies the post-selected channel
\begin{eqnarray}
|{\rm in}\rangle |0\rangle \mapsto 
|{\rm out}\rangle |0\rangle,
\end{eqnarray}
with
\begin{eqnarray}
|{\rm out}\rangle = \frac{e^{-\frac{\phi^2}{8} \, e_{11} } \, R_x(\theta) \,  |{\rm in}\rangle }{\| e^{-\frac{\phi^2}{8} \, e_{11} } \, R_x(\theta) \,  |{\rm in}\rangle \| } 
= \frac{ A_\tau \, |{\rm in}\rangle }{\| A_\tau \,  |{\rm in}\rangle \| } ,
\end{eqnarray}
where the second equality assumes circuit parameters
\begin{eqnarray}
\theta = \omega \tau \ {\rm and} \  \phi = \sqrt{ 8 \Gamma \tau}.
\end{eqnarray}
With these parameters, the circuit simulates (\ref{trotter step})
with nonnegative $\phi$.

The post-selection succeeds with probability 
\begin{eqnarray}
p = \| e^{-\frac{\phi^2}{8} \, e_{11} } \, R_x(\theta) \,  |{\rm in}\rangle \|^2.
\end{eqnarray}
However this is for one Trotter step. The probability of observing $K$ successful post-selections is $P = p_1 p_2 \cdots p_K$, where $p_k$ is the success probability for step $k$. Then $P \le (\max_k p_k)^K$, which is exponentially small in $K$ (assuming $\theta , \phi \neq 0 $). This makes the EPs $\omega = \pm \Gamma$ especially difficult to simulate because  both $\theta$ and $\phi$ have to be small, requiring more Trotter steps. These limitations are bypassed using a variational technique explained below.

\section{\label{sec:methods}Methods }

\subsection{\label{sec:machinelearning}Machine learning method}

\begin{figure*}
    \centering
    \includegraphics[width=2\columnwidth]{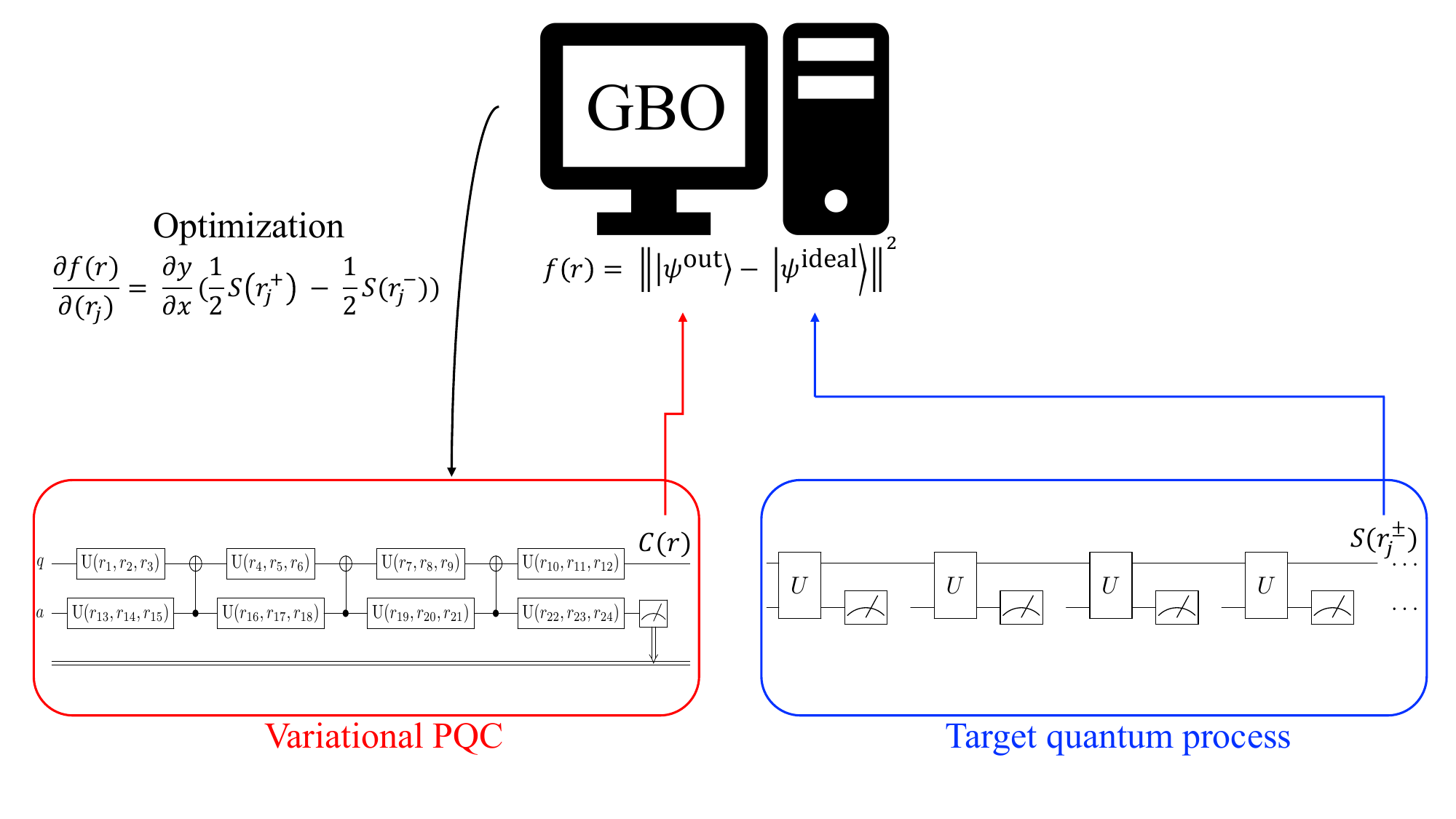}
    \caption{Scheme of the process to create the variational PQCs. A density matrix-based simulator simulates the target quantum process (outlined in Section \ref{sec:introduction}) shown within a blue box. The variational PQC ansatz (boxed in red) is similarly simulated with random initial parameters. The outcomes of both circuits are used to calculate the cost function $f(r)$ for the gradient-based optimization (shown in black). The parameters of the variational PQC are updated using the gradient, and the process of recalculating the output of the variational PQC and subsequent optimization is repeated as needed until a target fidelity is reached.}\label{fig:machine_learning_pipeline}
\end{figure*}

New developments in variational quantum circuits and machine learning models for non-unitary operations have provided an opportunity to develop short-depth quantum circuits capable of mirroring the same evolutionary dynamics of high-depth post-selected circuits. 

Being able to convert the post-selected circuit that simulates $H_{\text{eff}}$ into a constant and relatively shallow depth variational parameterized circuit would allow for a non-Hermitian Hamiltonian to be demonstrated on current quantum hardware and open the door for more complex quantum simulations. However, measurements and circuits that require post-selection, are non-unitary, destructive, and nonlinear. It follows that any parameterized quantum circuit (PQC) that seeks to approximate $H_{\text{eff}}$ must be capable of describing non-unitary quantum processes. 

In this study, we utilized the approach described in \cite{Xue2023} which presents a framework for describing non-unitary operators through variational parameterized quantum circuits. This method implements a supervised machine learning framework to output a parameterized circuit that stores all the information of our desired non-unitary quantum process in the parameters, denoted as $r$. Variational approaches are based on ansatzs, cost functions, and optimization methods. In the following subsections, we detail the specific roles and implementations of these core components.
In Fig.~\ref{fig:machine_learning_pipeline} we give a schematic of the machine learning approach.

\subsubsection{Ansatz}

\begin{figure*}
\centering
  \scalebox{1.1}{ 
    \Qcircuit @C=1em @R=1.0em @!R { \\
      \nghost{q } & \lstick{q  } & \gate{\mathrm{U}(r_{1},r_{2},r_{3})} & \targ & \gate{\mathrm{U}(r_{4},r_{5},r_{6})} & \targ & \gate{\mathrm{U}(r_{7},r_{8},r_{9})} & \targ & \gate{\mathrm{U}(r_{10},r_{11},r_{12})} & \qw & \qw & \qw\\
      \nghost{a } & \lstick{a  } & \gate{\mathrm{U}(r_{13},r_{14},r_{15})} & \ctrl{-1} & \gate{\mathrm{U}(r_{16},r_{17},r_{18})} & \ctrl{-1} & \gate{\mathrm{U}(r_{19},r_{20},r_{21})} & \ctrl{-1} & \gate{\mathrm{U}(r_{22},r_{23},r_{24})} & \meter\\
      \nghost{\mathrm{{}  }} & \lstick{\mathrm{{}  }} &  \cw & \cw & \cw & \cw & \cw & \cw & \cw & \dstick{_{_{\hspace{0.0em}0}}} \cw \ar @{<=} [-1,0] & \cw & \cw\\
    }
  }
  \caption{\label{ansatz} The variational parameterized quantum circuit ansatz. Using the ancilla qubit, $a$, we embed our desired non-unitary process within a unitary one.}
\end{figure*}
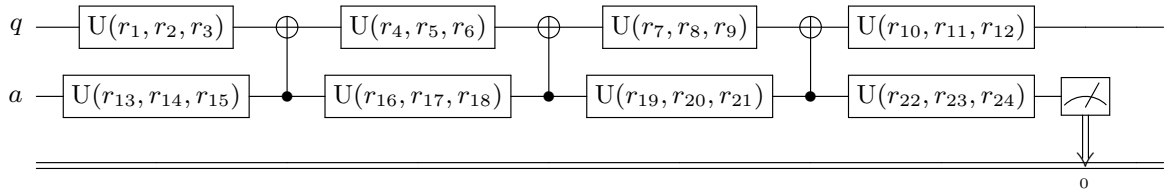

The ansatz is the shell of the circuit that will encode the parameters, $r$, and is what will be trained to minimize the cost function, $C(r)$. 
Fig. \ref{ansatz} depicts the ansatz that we used in this study. Using an ancilla in our ansatz, we embed our target non-unitary process within a unitary one, an approach known as dilation \cite{Dogra2021}, allowing us to do unitary machine learning. We use only a single ancilla qubit to implement the minimum dilation as prescribed in \cite{Xue2023}.

\subsubsection{Cost function}

Using the results from \cite{Xue2022}, we define our cost function, $f(r)$, as the function that will evaluate the distance between our variational PQC output, $C(r)$, and the target quantum process. We define the  cost function as $f(r) = 
 \vert\vert \vert\psi^{\text{out}} \rangle -\vert\psi^{\text{ideal}} \rangle \vert\vert ^2 %
 \label{costfxn1} = 2 - 2\operatorname{Re}\langle \psi_{j}^{\text{ideal}} \vert \psi_{j}^{\text{out}} \rangle $, where $ \vert\psi^{\rm ideal} \rangle $ and $\vert\psi^{\rm out} \rangle$ are the resulting states from the target quantum process and variational PQC respectively. 

\subsubsection{Gradient-based optimization}

Gradient based-optimization is one of the most practical optimization algorithms \cite{kingma2017adam}. The gradient of the cost function is calculated using the chain rule:
\begin{equation}
    \frac{\partial f(r)}{\partial r_j } = \frac{\partial f(r)}{\partial S(r) } \frac{\partial S(r)}{\partial r_j} = \frac{\partial f(r)}{\partial S(r)} \bigg( \frac{1}{2} S(r_j ^+) - \frac{1}{2} S(r_j^-) \bigg),
\label{gradient}
\end{equation} 
where $S$ is the output of the variational PQC and computation of $\frac{\partial S(r)}{\partial r_j}$ is done via the parameter-shift rule \cite{Wierichs_2022}. Here $r_j$ is the j-th parameter in the parameter list $r$, and $r_j^\pm = r_j \pm \frac{\pi}{2}$. The optimization process begins by assigning the $r_j$'s random initial values. These parameters correspond to the angles that parameterize the  $U$ gates of the ansatz circuit of Fig. \ref{ansatz}. The  $U$ gates in IBM's Qiskit framework represent the most general single-qubit rotation, allowing full control over a qubit's state on the Bloch sphere by parameterizing rotations with three angles. The corresponding matrix unitary of an arbitrary $U$ gate with rotation angles $r_a, r_b,$ and $r_c$ is 
\begin{equation}
U(r_a,r_b,r_c) = 
    \begin{pmatrix}
        \text{cos}(\frac{r_a}{2}) & -e^{ir_c}\text{sin}(\frac{r_a}{2}) \\
        e^{ir_b}\text{sin}(\frac{r_a}{2}) & e^{i(r_b + r_c)}\text{cos}(\frac{r_a}{2})
    \end{pmatrix}.
\end{equation}  
 $U$  gates allow for the parameter space to be large and allow for the necessary flexibility in the optimization process. By computing all terms in the cost function deterministically via a classical density matrix simulator and then using those results in the GBO, our optimization procedure is free of stochasticity from batch sampling and probabilistic outcomes.

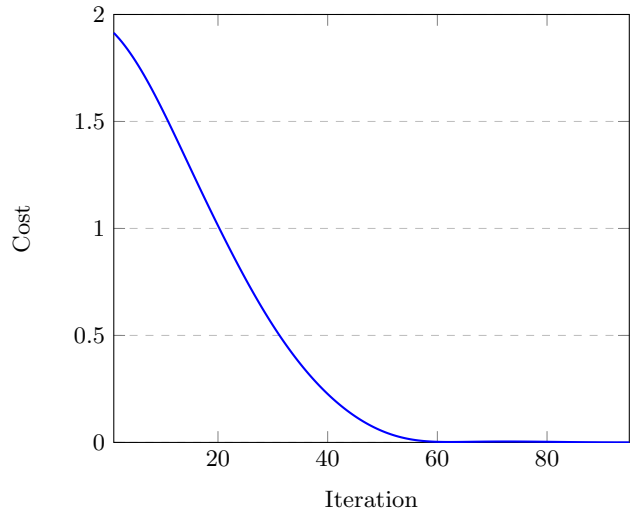
\begin{figure}
\centering
\begin{tikzpicture}
\begin{axis}[
    xlabel={Iteration},
    ylabel={Cost},
    xmin=1, xmax=95,
    ymin=0, ymax=2,
    legend pos=north east,
    ymajorgrids=true,
    grid style=dashed,
]

\addplot[
    color=blue,
    mark=none,
    smooth, 
    thick,
] coordinates {
    (1,1.91429063800899) (2,1.8861759774429214) (3,1.8544266722078306) (4,1.8191862415710842) (5,1.7806480295139981) (6,1.7390400306756169) (7,1.6946275873413033) (8,1.6477148754955413) (9,1.5986428955835852) (10,1.5477818896288336) (11,1.4955160032631047) (12,1.4422209320828587) (13,1.3882439163901001) (14,1.333899427082907) (15,1.2794749062952033) (16,1.225225689199801) (17,1.1713582251814307) (18,1.1180188611321846) (19,1.0652997518169447) (20,1.0132600376862517) (21,0.9619515140967927) (22,0.9114378679139483) (23,0.8618026704547963) (24,0.8131474295945571) (25,0.7655837522127118) (26,0.7192235711067301) (27,0.6741701470057515) (28,0.6305113620282896) (29,0.5883160252969455) (30,0.5476333154377093) (31,0.5084948592372318) (32,0.4709183766277436) (33,0.43491161925130317) (34,0.400475629543549) (35,0.3676069130865345) (36,0.3362985897016706) (37,0.30654081582966364) (38,0.2783208218845379) (39,0.25162288641632213) (40,0.22642850574637396) (41,0.20271690847337398) (42,0.18046592430974417) (43,0.15965308660131972) (44,0.14025676427319933) (45,0.12225709326689693) (46,0.10563649777886064) (47,0.09037963789964754) (48,0.07647267895969478) (49,0.06390184604222737) (50,0.05265130777975746) (51,0.04270052765280408) (52,0.03402132142454706) (53,0.02657494999446608) (54,0.020309636721488333) (55,0.015158907246401787) (56,0.011041095255360789) (57,0.007860237894549149) (58,0.005508411268927427) (59,0.0038693543377437134) (60,0.0028230336995920258) (61,0.002250651632233147) (62,0.002039529298791809) (63,0.0020873241517713748) (64,0.0023051599286794122) (65,0.0026194311629583744) (66,0.002972249783387504) (67,0.0033206864131773983) (68,0.003635091868781304) (69,0.0038968507580836587) (70,0.00409592117007862) (71,0.004228465860967923) (72,0.0042947998882605365) (73,0.00429778620815191) (74,0.004241720183492337) (75,0.004131667776365644) (76,0.003973167500270991) (77,0.003772176035071295) (78,0.003535131351585896) (79,0.003269021940117245) (80,0.0029813807033256534) (81,0.002680160214992622) (82,0.0023734849012451154) (83,0.0020693085557663693) (84,0.0017750275481485112) (85,0.0014971087327944055) (86,0.0012407867340407108) (87,0.0010098706575001337) (88,0.0008066796288674905) (89,0.0006321047234090216) (90,0.00048577618717948745) (91,0.0003663024458291364) (92,0.0002715425999062493) (93,0.0001988765288694072) (94,0.00014544461197685266) (95,0.00010833990659664394)
};

\end{axis}
\end{tikzpicture}
\caption{The gradient-based optimization algorithm’s rapid minimization of the cost function, transitioning from an initial peak near 2 to a nadir value near 0, illustrating its robust capability in navigating complex loss landscapes toward the global minimum.}
\label{gradient_minimization}
\end{figure}

Fig. \ref{gradient_minimization} presents an example of rapid cost minimization and is an excellent example of the robustness of our GBO method.  Even when the initial guess for the initial $r_j$'s results in the worst-case scenario of an ansatz with the maximum possible value of the cost function $f(r)$ of two, our optimization method successfully yields the $r_j$'s that correspond to a cost near zero.

\section{\label{sec:results}Results}

\begin{figure}
    \centering
    \begin{tikzpicture}
    \begin{axis}[
      width=\columnwidth,
      height=6cm,
      axis y line*=left,
      ymin=0, ymax=1,
      xlabel=Time steps,
      ylabel=$P_0$,
      ylabel near ticks,
      ylabel style={shift={(-5pt,0pt)}},
      legend style={at={(0.5,1.03)}, anchor=south, legend columns=2},
      legend entries={$\omega = 8\Gamma$ (Simulation) ,$\omega = 0.89\Gamma$ (Simulation), $\omega = 8\Gamma$ (IBM Brisbane), $\omega = 0.89\Gamma$ (IBM Brisbane)}
    ]
    \addplot[smooth,mark=o,blue] 
      coordinates{
        (1,0.999939999399994)
        (50,0.9412394123941239)
        (100,0.7894552618683308)
        (150,0.5935759357593576)
        (200,0.3780937809378094)
        (250,0.17713062783767025)
        (300,0.04033161326453058)
        (396,0.07451149022980459)
        (454,0.3108524340973639)
        (499,0.5622)
        (552,0.8312681887732142)
        (597,0.98536)
        (663,0.9780895617912359)
        (693,0.91684)
        (788,0.5752615052301046)
    };
    \addplot[smooth,mark=o,red] 
      coordinates{
        (0,0.999989999299951)
        (2,0.9999199959998)
        (50,0.963)
        (103,0.90532)
        (151,0.8619472389447789)
        (195,0.8317983179831798)
        (254,0.8064222568902756)
        (302,0.7878036341090233)
        (340,0.77329)
        (392,0.76492)
        (444,0.754045080901618)
        (514,0.7483699347973919)
        (599,0.7459371343420907)
        (709,0.7396943816628998)
        (777,0.7381538153815381)
    };

    \addplot[smooth,mark=*,blue] 
      coordinates{
        (1, 0.9226863484087102)
        (50, 0.8673020527859238)
        (100, 0.74497448182481)
        (150, 0.5520974997384663)
        (200, 0.35544626974992155)
        (250, 0.17172295363480253)
        (300, 0.06120644825793032)
        (396, 0.08709191882493382)
        (454, 0.2945331032995471)
        (499, 0.5322674881818664)
        (552, 0.764257135236063)
        (597, 0.9326454130088814)
        (663, 0.8955863149647427)
        (693, 0.8435550935550935)
        (788, 0.5221553753596652)
    };

    \addplot[smooth,mark=*,red] 
      coordinates{
(0, 0.9470744953376371)
(2, 0.9499078057775046)
(50, 0.9078650533223954)
(103, 0.84375)
(151, 0.8013599835153513)
(195, 0.7743789776226647)
(254,0.7425031236984589)
(302,0.7135157545605307)
(340,0.7129275601942751)
(392,0.7010374096196165)
(444,0.691993078496146)
(514,0.6949196207749382)
(599, 0.7200738537285876)
(709, 0.7135400675468222)
(777, 0.7163957126006462)
    };
    \end{axis}
    \end{tikzpicture}
  \caption{\label{probexp_results}{Probability of the system qubit remaining in the \(|0\rangle\) state after $t$ simulated time steps as a function of $\Gamma$, with $\omega=0.01$ fixed. Oscillations in \(P_0\) when $\omega = 8 \Gamma $ 
 (blue curves) indicate the system is in a weakly dissipative phase, far from the EP in the $PT$ symmetric regime. The absence of oscillations when $\omega = 0.89 \Gamma $  (red curves) signifies the system's dissipative dynamics close to the EP in the broken symmetry regime. Results from IBM Brisbane \cite{BrisbaneCharacteristics2024} and classical simulators are both shown.}}
    \label{fig:pgfplot-two}
\end{figure}
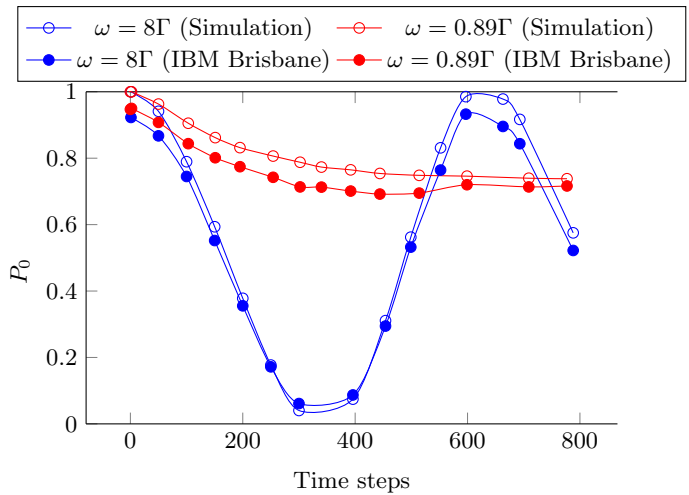

Due to the limitations discussed in \ref{sec:nonhermsim}, simulating non-Hermitian Hamiltonians has not yet been demonstrated. Our approach overcomes these limitations, as we now explain. In the following simulations, we choose $\omega = 0.01$ and vary $\Gamma$.

\subsection{\label{sec:[probdynam]}Probability dynamics}

After initializing the system qubit to the state \(|0\rangle\) and evolving it using the variational PQC, we then measure the $|0\rangle$ state probability $P_0 = |\langle 0 | \psi \rangle|^2$ versus time. The resulting dynamics demonstrate pronounced non-Hermitian effects. 
As discussed above, the EP marks a phase transition from a $PT$ symmetric to a $PT$ broken phase. Thus, we perform probability dynamics simulations by choosing parameters to access the parameter space close to the EP. The results of the probability dynamics studys are shown in Fig. \ref{probexp_results}. The observations are in good agreement with theoretical predictions of \cite{Zhang2021}, validating the model's efficacy in describing $H_{\text{eff}}$. Details on the layout and characteristics of the qubits used at the time of these demonstrations can be found at \cite{BrisbaneCharacteristics2024}.

\subsection{\label{sec:level4}Expectation of $\sigma_z$ }

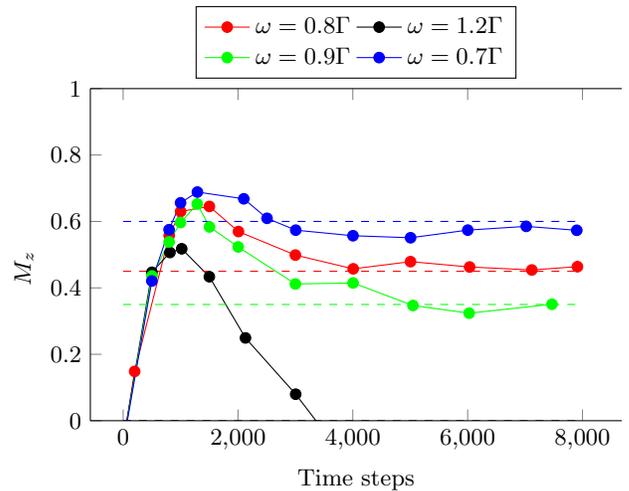
\begin{figure}
    \centering
    \begin{tikzpicture}
    
    \begin{axis}[
      width=\linewidth,
      height=6cm,
      axis y line*=left,
      ymin=0, ymax=1,
      xlabel=Time steps,
      ylabel=$M_z$,
      ylabel near ticks,
      ylabel style={shift={(-5pt,0pt)}},
      legend style={at={(0.5,1.03)}, anchor=south, legend columns=2},
      legend entries={$\omega = 0.8\Gamma$, $\omega = 1.2\Gamma$, $\omega = 0.9\Gamma$,  $\omega = 0.7\Gamma$}
    ]
\addplot[red, mark=*, mark size=2pt] coordinates {
(19, -0.05234555242245581)
(200, 0.14844471820141675)
(802, 0.5578101663490755)
(1001, 0.6300229767679346)
(1502, 0.6452223757612978)
(2008, 0.5695905238435617)
(3000, 0.4987714987714988)
(4007, 0.45732853536709817)
(5005, 0.4789907149976394)
(6034, 0.46288346769971733)
(7117, 0.45391545391545396)
(7912, 0.4638239484933242)
};


\addplot[black, mark=*, mark size=2pt] coordinates {
(15, -0.04898172323759792)
(501, 0.44687467606509795)
(815, 0.5065339141257001)
(1020, 0.5176660925838943)
(1501, 0.43357557379620437)
(2130, 0.2493116525533794)
(3005, 0.07979448463877525)
(3400, -0.009382043084019076)
(3703, -0.03281160626072488)
(5179, -0.047087403731844724)
};

\addplot[green, mark=*, mark size=2pt] coordinates {
(18, -0.05449435278197057)
(502, 0.4360399300195534)
(801, 0.5383668444809191)
(1000, 0.5967120472643206)
(1290, 0.6525065021163752)
(1504, 0.5836271654146865)
(2003, 0.5232742322824588)
(3000, 0.4117157856437445)
(4004, 0.4148626144879268)
(5047, 0.3469560166851471)
(6026, 0.3238205560235889)
(7469, 0.3506300963676798)
};

\addplot[blue, mark=*, mark size=2pt] coordinates {
(17, -0.05464224205417256)
(500, 0.42071097372488403)
(800, 0.5752439650744736)
(1000, 0.6560627017058551)
(1293, 0.6885538179410116)
(2101, 0.6681302971510263)
(2506, 0.609462013998876)
(3006, 0.5737670732978835)
(4003, 0.5570077568802465)
(5006, 0.5509032943676939)
(6002, 0.5739162265334936)
(7020, 0.585189075630252)
(7900, 0.5733750978856695)
};

\draw[red, dashed] (axis cs:0,0.45) -- (axis cs:8000,0.45);
\draw[blue, dashed] (axis cs:0,0.6) -- (axis cs:8000,0.6);
\draw[black, dashed] (axis cs:0,0) -- (axis cs:8000,0);
\draw[green, dashed] (axis cs:0,0.35) -- (axis cs:8000,0.35);
    \end{axis}
    \end{tikzpicture}
  \caption{\label{mz_results}{This plot shows the dynamic response of \(M_{z}\) verses time, derived from measurements on the IBM Brisbane platform. 
  We observe the convergence of the magnetization to a stationary state when \(\Gamma > \omega\), mirroring the same behavior when run on the simulator (see Fig.~\ref{fig:pgfplot-two}). The trend reveals that as we approach the EP, more steps are needed to reach a steady state. The dashed horizontal lines are theoretical results from \cite{Zhang2021}.}}
    \label{fig:pgfplot-two}
\end{figure}

\begin{figure}
    \centering
    \begin{tikzpicture}
    
    \begin{axis}[
      width=\columnwidth,
      height=6cm,
      axis y line*=left,
      ymin=0, ymax=0.8,
      xlabel=$\omega / \Gamma$,
      ylabel=$M_z$,
      ylabel near ticks,
      ylabel style={shift={(-5pt,0pt)}},
      legend style={at={(0.5,1.03)}, anchor=south, legend columns=2},
      legend entries={Simulator, IBM Brisbane}
    ]

    \addplot[blue, mark=o, mark size=2pt] coordinates {
        (0.7, 0.7)
        (0.8, 0.6)
        (0.9, 0.45)
        (1, 0)
        (1.1, 0)
        (1.2,0)
    };
        \addplot[blue, mark=*, mark size=2pt] coordinates {
        (0.7, 0.6)
        (0.8, 0.45)
        (0.9, 0.35)
        (1,0)
        (1.1, 0)
        (1.2, 0)
    };
    \addplot[red, mark=x, only marks, mark size=4pt] coordinates {
        (1,0)
    };
\node[above right] at (axis cs:1,0) {EP};

    \end{axis}
\end{tikzpicture}

\caption{\label{mx_phasetransition} This graph shows the asymptotic value of $M_z$ as a function of $\omega/ \Gamma$, with the EP existing at $\omega = \Gamma$. The phase transition is apparent as we sample around the EP. 
The regime $\omega/\Gamma < 1$ is the broken $PT$ symmetry phase. Classical simulation results and results from IBM Brisbane are both shown.}
\end{figure}
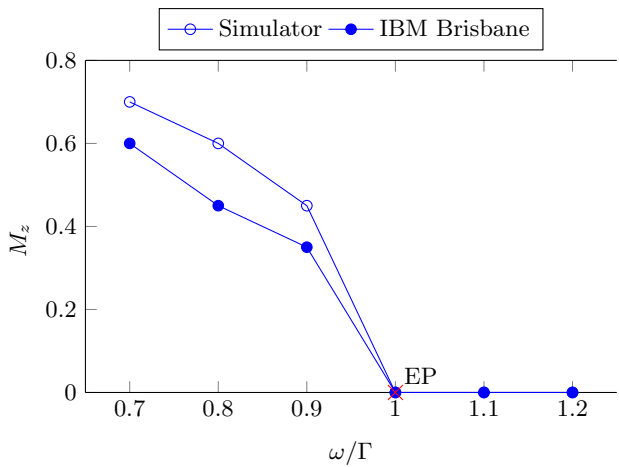

This demonstration measures the expectation value of the Pauli observable \(M_z \equiv \langle \sigma_z \rangle\), which can be interpreted as a magnetization. Beginning with the system qubit in the fully mixed state, $ \rho = I/2$, prepared by entanglement with an additional qubit (not shown in Fig.~\ref{fig:model}), we study the dynamics of \(M_z\) to observe a phase transition and convergence to stationary values of \(M_z\) around the EP. Fig. \ref{mz_results} shows the measured data, demonstrating convergence to stable values in the broken symmetry regime where \( \Gamma > \omega \), after  extensive evolution surpassing 4000 time steps. The plot shows $M_z$ undergoing a phase transition at the EP. To see this, Fig.~\ref{mx_phasetransition} plots the asymptotic magnetization values versus $\omega$, for fixed $\Gamma$. 
This plot shows that $M_z$ at long times acts as an order parameter indicating the breaking of $PT$ symmetry.

\begin{figure}
    \begin{tikzpicture}
    \begin{axis}[
      width=\linewidth,
      height=6cm,
      axis y line*=left,
      ymin=0, ymax=1,
      xlabel=Time steps,
      ylabel=$M_z$,
      ylabel near ticks,
      ylabel style={shift={(-5pt,0pt)}},
      legend style={at={(0.5,1.03)}, anchor=south, legend columns=2},
      legend entries={$\omega = 0.8\Gamma$, $\omega = 1.2\Gamma$, $\omega = 0.9\Gamma$,  $\omega = 0.7\Gamma$}
    ]

\addplot[red, mark=o, mark size=2pt] coordinates {
(19, 0.021389856721785953)
(200, 0.2522293165641627)
(802, 0.7053946156488826)
(1001, 0.7480263584290856)
(1502, 0.7717050908441531)
(2008, 0.739207852735404)
(3000, 0.669972859988969)
(4007, 0.6336479133560775)
(5005, 0.6136416107923819)
(6034, 0.6070317051540843)
(7117, 0.6022678850785501)
(7912, 0.6108198240827877)
};

\addplot[black, mark=o, mark size=2pt] coordinates {
(1, 0.006477535469917763)
(15, 0.024601368600658518)
(501, 0.5144474991058551)
(815, 0.6240761850911014)
(1020, 0.6360985683123224)
(1501, 0.5576575424092575)
(2130, 0.3881611076354502)
(3005, 0.17842761327502005)
(3400, 0.0787244528490843)
(3703, 0.009695322750159219)
(5179, 0)
(7912, 0)
};

\addplot[green, mark=o, mark size=2pt] coordinates {
(18, 0.028475810703143613)
(502, 0.5353967944854164)
(801, 0.6809107720852022)
(1000, 0.7248170610603996)
(1504, 0.7284310173015067)
(2003, 0.6745002567326495)
(3000, 0.5729124733282872)
(4004, 0.5119761451652413)
(5047, 0.4788553335018799)
(6026, 0.46064311919530976)
(7469, 0.4643162501047073)
};
\addplot[blue, mark=o, mark size=2pt] coordinates {
(17, 0.027185350353330365)
(500, 0.5442182752440926)
(800, 0.7111859145401692)
(1000, 0.7714106152343918)
(1293, 0.8085544758429274)
(2101, 0.7937167806045626)
(2506, 0.772661922444189)
(3006, 0.7546252423337029)
(4003, 0.7295596579569075)
(5006, 0.7197742869472354)
(6002, 0.7151873012909302)
(7020, 0.7132595029155975)
(7900, 0.7125491454076842)
};


\draw[red, dashed] (axis cs:0,0.6) -- (axis cs:8000,0.6);
\draw[blue, dashed] (axis cs:0,0.7) -- (axis cs:8000,0.7);
\draw[black, dashed] (axis cs:0,0) -- (axis cs:8000,0);
\draw[green, dashed] (axis cs:0,0.45) -- (axis cs:8000,0.45);
    \end{axis}
    \end{tikzpicture}
  \caption{\label{mz_results_simulate}{
 Same as Fig.~\ref{mz_results} but obtained by classical simulation.}} 
 \label{fig:pgfplot-two}
\end{figure}
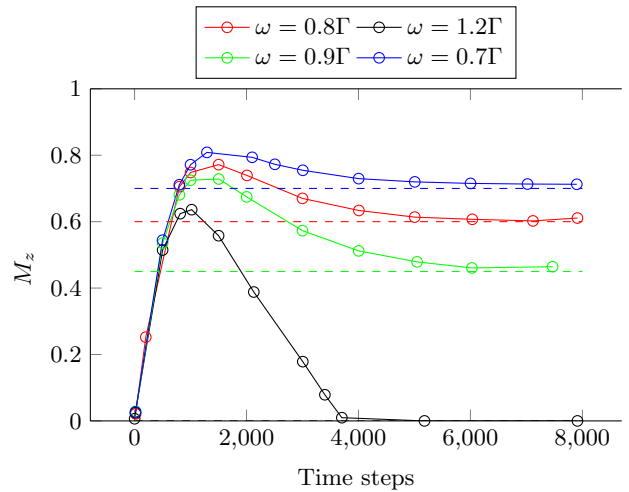

\subsection{\label{sec:level5} Entanglement monotonicity violation and state evolution near EP}

Finally,  using classical simulations (not measured data), we explore the distinctively non-Hermitian phenomenon of entanglement monotonicity violation \cite{PhysRevA.90.054301, PhysRevLett.119.190401} as a qubit evolves under $H_{\rm eff}$. We analyze the behavior near the EP, where we observe an entanglement that {\it increases} under a local operation in the $PT$-symmetric phase. We begin with the system qubit $q$ in a Bell state with a second qubit, $q'$. The qubit $q'$ remains undisturbed throughout the process (no gates are applied). While $q$ undergoes local $ H_{\rm eff} $ evolution, we measure the entanglement between the non-Hermitian qubit $q$ and its entangled partner $q'$. 
In this work we use concurrence as the measure of entanglement in a two-qubit state $\rho \in {\mathbb C}^{4 \times 4}$, through the matrix 
\begin{eqnarray}
A :=  \, \rho {\tilde \rho} , \ \  
{\tilde \rho} := 
\sigma^y \! \otimes \! \sigma^y \,  {\bar \rho} \,\sigma^y \! \otimes \! \sigma^y ,
 \label{t matrix}
\end{eqnarray}
where ${\bar \rho} $ is the complex conjugate of $\rho$ in the standard ($\sigma^z$ eigenstate) basis \cite{PhysRevLett.80.2245}. Concurrence is defined through the eigenvalues of $A$, which are nonnegative with square roots $\lambda_i$:
\begin{eqnarray}
{\rm spec}(A) = \{ \lambda^2_1, \lambda^2_2, \lambda^2_3,   \lambda^2_4 \} , \ \ 
 \lambda_1 \ge  \lambda_2 \ge  \lambda_3 \ge  \lambda_4 \ge 0. \ \ \ \ \ \
 \label{t spectrum}
\end{eqnarray}
Then 
\begin{eqnarray}
C(\rho) = \max \{  \Delta , 0 \} ,
\ \  \Delta := \lambda_1 - \lambda_2 -  \lambda_3 -  \lambda_4 .
 \label{tangle definiition}
\end{eqnarray}

\begin{figure}
\centering
\begin{tikzpicture}
    \begin{axis}[
      width=\columnwidth,
      height=6cm,
      axis y line*=left,
      ymin=0, ymax=1.0,
      xlabel=Time steps,
      ylabel=$C_{q, q^{'}}$,
      ylabel near ticks,
      ylabel style={shift={(-5pt,0pt)}},
    legend style={at={(0.5,1.03)}, anchor=south, legend columns=-1},
      legend entries={$\omega = 0.89\Gamma$,$\omega=2\Gamma$, $\omega=8\Gamma$}
    ]

    \addplot[red, mark=none] coordinates {
(1.0,0.9999999937348335)
(2.0,0.9999362347097944)
(3.0,0.9997450152681394)
(4.0,0.9994264618679473)
(5.0,0.9989808395967905)
(6.0,0.9984084295481498)
(7.0,0.9977096949194179)
(8.0,0.9968851276992975)
(9.0,0.9959353361617912)
(10.0,0.9948610104515189)
(11.0,0.9936629359372424)
(12.0,0.9923419751776371)
(13.0,0.9908990916822421)
(14.0,0.9893353173577752)
(15.0,0.9876517752217677)
(16.0,0.9858496617532942)
(17.0,0.9839302563485126)
(18.0,0.9818949291664201)
(20.0,0.9774822419760217)
(21.0,0.9751079569312471)
(23.0,0.9700316631576188)
(24.0,0.9673330884806477)
(28.0,0.9555123706067488)
(29.0,0.9523102972114569)
(30.0,0.9490135023426823)
(31.0,0.9456240777829606)
(33.0,0.9385759518297331)
(34.0,0.934921606031955)
(35.0,0.9311833688538842)
(36.0,0.927363447820039)
(37.0,0.9234641717679142)
(38.0,0.9194877761071133)
(39.0,0.915436591054059)
(40.0,0.9113129062490432)
(41.0,0.9071190672559516)
(42.0,0.902857379181688)
(44.0,0.8941398540842526)
(45.0,0.8896886776095939)
(47.0,0.880613122581779)
(48.0,0.8759933974163915)
(50.0,0.8666015171531386)
(51.0,0.8618339333890883)
(53.0,0.8521667937487528)
(54.0,0.8472716878127461)
(56.0,0.837369348980132)
(57.0,0.8323664413368633)
(58.0,0.827331867714171)
(59.0,0.8222677221238381)
(60.0,0.8171760475378637)
(62.0,0.8069182367158189)
(63.0,0.8017560410970842)
(64.0,0.7965742308925828)
(65.0,0.7913747081673718)
(67.0,0.7809298609168426)
(68.0,0.775688142223788)
(69.0,0.7704358966268087)
(70.0,0.7651748291409898)
(72.0,0.7546328444369166)
(74.0,0.7440750474935895)
(75.0,0.7387940612222633)
(78.0,0.7229602646582183)
(79.0,0.7176900140266365)
(80.0,0.7124258270471987)
(83.0,0.6966821201917882)
(86.0,0.6810363659794568)
(87.0,0.6758480084029793)
(90.0,0.660377705471838)
(91.0,0.6552556367696875)
(99.0,0.6150219485232992)
(100.0,0.6100963262905374)
(101.0,0.6051958236382335)
(103.0,0.595472334309051)
(105.0,0.5858555419674852)
(106.0,0.5810883088713854)
(107.0,0.5763490788078309)
(111.0,0.5576796609481961)
(112.0,0.5530858440384395)
(113.0,0.548522034724051)
(114.0,0.5439884842616525)
(115.0,0.5394854196717269)
(117.0,0.5305716240001528)
(119.0,0.5217821611823517)
(120.0,0.5174344462002292)
(121.0,0.513118253845991)
(122.0,0.5088336972202908)
(124.0,0.5003598298210035)
(127.0,0.48788818311343796)
(128.0,0.4837949169841804)
(130.0,0.4757043787215574)
(131.0,0.47170710953617623)
(133.0,0.46380846136920706)
(134.0,0.4599069971100869)
(135.0,0.4560373847035567)
(138.0,0.4446189489309133)
(141.0,0.43348437892788855)
(148.0,0.4085909780197534)
(149.0,0.4051570941740169)
(150.0,0.40175333979706584)
(151.0,0.398379553399393)
(152.0,0.39503558001677813)
(154.0,0.38843635715211355)
(159.0,0.3724475890621856)
(160.0,0.3693356577843881)
(162.0,0.36319608734540065)
(163.0,0.3601680487480177)
(164.0,0.3571675803232272)
(168.0,0.3454373385367875)
(169.0,0.3425716600411451)
(171.0,0.33691911620531534)
(176.0,0.3232377453819686)
(177.0,0.3205769289299134)
(184.0,0.30262995659417025)
(185.0,0.3001605446971267)
(196.0,0.27446861856015725)
(198.0,0.2700748811286481)
(199.0,0.26790863859536734)
(202.0,0.26153006337157364)
(204.0,0.2573758858436364)
(205.0,0.25532768649006)
(209.0,0.24732313554172236)
(211.0,0.24343123233990654)
(217.0,0.23217978434567987)
(218.0,0.23036457932234874)
(219.0,0.2285660747847725)
(220.0,0.2267840963998197)
(225.0,0.21811637678658655)
(228.0,0.21310367551567136)
(231.0,0.20822674378926506)
(232.0,0.20663056147473613)
(235.0,0.20192830575516457)
(241.0,0.19289915898145754)
(244.0,0.18856472374813044)
(245.0,0.18714570207069958)
(248.0,0.18296416283959968)
(250.0,0.18023813258631197)
(252.0,0.17756027644111175)
(260.0,0.16731129089909091)
(262.0,0.1648600232410457)
(263.0,0.1636504221161674)
(267.0,0.1589163938924014)
(268.0,0.1577584576207595)
(269.0,0.15661053846567166)
(271.0,0.15434435859786766)
(276.0,0.14884734629032784)
(277.0,0.14777604666308616)
(278.0,0.14671390021325656)
(280.0,0.14461668115068074)
(281.0,0.14358144267723066)
(286.0,0.1385353058812889)
(290.0,0.13464941411275538)
(291.0,0.13369822947052107)
(292.0,0.13275499337355906)
(294.0,0.13089207313363543)
(299.0,0.12636859941493836)
(300.0,0.12548624833019525)
(307.0,0.11950964605334001)
(308.0,0.11868356900853325)
(313.0,0.11465287950245796)
(324.0,0.10633924621253299)
(325.0,0.10561922006396797)
(327.0,0.10419626922865352)
(331.0,0.10141725980609155)
(338.0,0.09676008588128462)
(346.0,0.09174035449223002)
(350.0,0.08934467278666053)
(359.0,0.08421541446316717)
(369.0,0.07891179770525389)
(389.0,0.06941806067069102)
(390.0,0.06897893779563984)
(395.0,0.06683047504983922)
(401.0,0.06435246001281673)
(402.0,0.06394973253876185)
(409.0,0.06120954877257779)
(415.0,0.05896661710878506)
(418.0,0.057880162901025954)
(421.0,0.05681630874184465)
(428.0,0.05441873927877294)
(431.0,0.05342628946751386)
(436.0,0.0518171001130128)
(454.0,0.04645796619986248)
(475.0,0.04097139274737202)
(476.0,0.04072865905728377)
(484.0,0.03884294323768008)
(495.0,0.0364052193122282)
(496.0,0.036192074710367166)
(500.0,0.035353096384511694)
(501.0,0.03514669687804678)
(520.0,0.031464925788220556)
(527.0,0.03021615286708902)
(538.0,0.028361020544215885)
(544.0,0.027401323842965748)
(585.0,0.02170818435851294)
(598.0,0.020178259924470317)
(610.0,0.018867346867699194)
(614.0,0.018450677287938193)
(681.0,0.012740767189855007)
(698.0,0.01160947933427306)
(1000,0)
    };
    
\addplot[black, mark=none] coordinates{
( 1 , 0.9999999937348335 )
( 2 , 0.9999874517180264 )
( 3 , 0.9999498291696298 )
( 4 , 0.9998871418050057 )
( 5 , 0.9997993854340214 )
( 6 , 0.9996865865099154 )
( 7 , 0.9995487899211162 )
( 8 , 0.9993860104506477 )
( 9 , 0.999198278386677 )
( 10 , 0.9989856478912668 )
( 11 , 0.9987481472110843 )
( 12 , 0.9984858573979088 )
( 13 , 0.9981988369572943 )
( 14 , 0.9978871239970242 )
( 15 , 0.997550803049135 )
( 16 , 0.9971899461744712 )
( 17 , 0.9968046393446922 )
( 18 , 0.9963949559457345 )
( 19 , 0.9959610026559118 )
( 20 , 0.9955028764446694 )
( 21 , 0.9950206605046104 )
( 22 , 0.9945144815404214 )
( 23 , 0.9939844361445337 )
( 24 , 0.9934306364810075 )
( 25 , 0.9928532395945305 )
( 26 , 0.9922523295044358 )
( 27 , 0.9916280526974619 )
( 28 , 0.9909805556678561 )
( 29 , 0.9903099851132637 )
( 30 , 0.9896164511833407 )
( 31 , 0.9889001315029199 )
( 32 , 0.9881611882278044 )
( 33 , 0.9873997502442845 )
( 34 , 0.9866160180039634 )
( 35 , 0.9858101177893227 )
( 36 , 0.9849822477789903 )
( 37 , 0.9841325907105836 )
( 38 , 0.9832613101745764 )
( 39 , 0.9823685844790491 )
( 40 , 0.9814546284174303 )
( 41 , 0.980519586547451 )
( 42 , 0.9795637210267675 )
( 43 , 0.9785871653310287 )
( 44 , 0.9775901637833739 )
( 45 , 0.9765729009955846 )
( 46 , 0.9755355923999617 )
( 47 , 0.9744784447857837 )
( 48 , 0.9734016725116617 )
( 49 , 0.9723054962492853 )
( 50 , 0.9711901644325155 )
( 51 , 0.9700558615549336 )
( 52 , 0.9689028273594718 )
( 53 , 0.9677312908028802 )
( 54 , 0.9665414706180803 )
( 55 , 0.9653336374072478 )
( 56 , 0.9641079815464109 )
( 57 , 0.9628647792427424 )
( 58 , 0.9616042506502478 )
( 59 , 0.960326653207765 )
( 60 , 0.9590322177138422 )
( 61 , 0.9577211852071938 )
( 62 , 0.9563938221516033 )
( 63 , 0.9550503522671048 )
( 64 , 0.9536910537474964 )
( 65 , 0.9523161836917458 )
( 66 , 0.9509259643555896 )
( 67 , 0.9495206819035954 )
( 68 , 0.9481005731835307 )
( 69 , 0.9466658957832631 )
( 70 , 0.9452169183482456 )
( 71 , 0.9437538971002997 )
( 72 , 0.9422770822911894 )
( 73 , 0.9407867505689537 )
( 74 , 0.9392831485737233 )
( 75 , 0.9377665510874581 )
( 76 , 0.9362372154742814 )
( 77 , 0.9346953883712529 )
( 78 , 0.9331413673655867 )
( 79 , 0.9315753694196265 )
( 80 , 0.9299977091283034 )
( 81 , 0.92840860999628 )
( 82 , 0.9268083349045648 )
( 83 , 0.9251971864447958 )
( 84 , 0.9235753929625399 )
( 85 , 0.9219432137991734 )
( 86 , 0.9203009311827395 )
( 87 , 0.9186487806740741 )
( 88 , 0.9169870530786044 )
( 89 , 0.915315985873706 )
( 90 , 0.9136358424671048 )
( 91 , 0.9119468941324308 )
( 92 , 0.9102493767222016 )
( 93 , 0.9085435576776233 )
( 94 , 0.9068297157098713 )
( 95 , 0.9051080671087894 )
( 96 , 0.9033788773537725 )
( 97 , 0.9016424014314486 )
( 98 , 0.8998989023518106 )
( 99 , 0.8981486305805929 )
( 100 , 0.8963918085439754 )
( 101 , 0.8946287173939993 )
( 102 , 0.8928595759862588 )
( 103 , 0.8910846402891656 )
( 104 , 0.889304144300337 )
( 105 , 0.887518340285408 )
( 106 , 0.885727461547886 )
( 107 , 0.8839317531746949 )
( 108 , 0.8821314391087005 )
( 109 , 0.8803267764742493 )
( 110 , 0.8785179661656956 )
( 111 , 0.8767052643390506 )
( 112 , 0.8748888987085485 )
( 113 , 0.8730690792513477 )
( 114 , 0.8712460420690212 )
( 115 , 0.8694200213469702 )
( 116 , 0.8675912276680697 )
( 117 , 0.8657598697642646 )
( 118 , 0.8639261815848597 )
( 119 , 0.8620903658760497 )
( 120 , 0.8602526345965155 )
( 121 , 0.8584132164157167 )
( 122 , 0.8565723172988736 )
( 123 , 0.854730129799491 )
( 124 , 0.8528868651109615 )
( 125 , 0.8510427078824427 )
( 126 , 0.8491978944376944 )
( 127 , 0.8473526019721109 )
( 128 , 0.845507019860862 )
( 129 , 0.8436613593896668 )
( 130 , 0.8418158006286565 )
( 131 , 0.8399705231985993 )
( 132 , 0.838125729126548 )
( 133 , 0.8362815977440948 )
( 134 , 0.8344383108127144 )
( 135 , 0.8325960512113545 )
( 136 , 0.8307549833320792 )
( 137 , 0.8289152947899978 )
( 138 , 0.8270771645676436 )
( 139 , 0.8252407313467647 )
( 140 , 0.8234062008027536 )
( 141 , 0.8215737227825847 )
( 142 , 0.8197434506708177 )
( 143 , 0.8179155693995338 )
( 144 , 0.8160902141625874 )
( 145 , 0.8142675391128272 )
( 146 , 0.8124477345311578 )
( 147 , 0.8106309130064361 )
( 148 , 0.8088172266413852 )
( 149 , 0.8070068588086785 )
( 150 , 0.8051999176257437 )
( 151 , 0.8033965569910473 )
( 152 , 0.8015969210102543 )
( 153 , 0.7998011518309645 )
( 154 , 0.7980093718884601 )
( 155 , 0.7962217222228916 )
( 156 , 0.7944383364547385 )
( 157 , 0.792659329883436 )
( 158 , 0.7908848477595731 )
( 159 , 0.7891150067974451 )
( 160 , 0.7873499313701174 )
( 161 , 0.7855897436119827 )
( 162 , 0.7838345707350995 )
( 163 , 0.7820845019585564 )
( 164 , 0.7803396656855225 )
( 165 , 0.7786001795763638 )
( 166 , 0.776866150673408 )
( 167 , 0.7751376851561793 )
( 168 , 0.7734148977904607 )
( 169 , 0.7716978705455411 )
( 170 , 0.7699867235577229 )
( 171 , 0.7682815521447249 )
( 172 , 0.7665824567122129 )
( 173 , 0.764889526101752 )
( 174 , 0.7632028673318183 )
( 175 , 0.7615225512382088 )
( 176 , 0.7598486825916656 )
( 177 , 0.7581813571778353 )
( 178 , 0.7565206499917404 )
( 179 , 0.7548666442266262 )
( 180 , 0.7532194238555089 )
( 181 , 0.7515790768651659 )
( 182 , 0.7499456788710945 )
( 183 , 0.7483193056343969 )
( 184 , 0.7467000366000344 )
( 185 , 0.7450879417326849 )
( 186 , 0.7434830829632901 )
( 187 , 0.741885554483907 )
( 188 , 0.7402954124030775 )
( 189 , 0.7387127265442568 )
( 190 , 0.7371375666503612 )
( 191 , 0.7355699903381423 )
( 192 , 0.7340100556146046 )
( 193 , 0.7324578378127367 )
( 194 , 0.7309133866149677 )
( 195 , 0.7293767660160395 )
( 196 , 0.7278480280977477 )
( 197 , 0.7263272279228528 )
( 198 , 0.7248144293333968 )
( 199 , 0.7233096741530528 )
( 200 , 0.721813008440146 )
( 201 , 0.7203244971281141 )
( 202 , 0.718844182118454 )
( 203 , 0.717372098147925 )
( 204 , 0.7159083138333783 )
( 205 , 0.7144528555076612 )
( 206 , 0.7130057711280045 )
( 207 , 0.7115671144503137 )
( 208 , 0.7101369085418658 )
( 209 , 0.7087151989755298 )
( 210 , 0.7073020398727706 )
( 211 , 0.7058974400585255 )
( 212 , 0.7045014547088271 )
( 213 , 0.7031141120305652 )
( 214 , 0.7017354516044352 )
( 215 , 0.7003654964286206 )
( 216 , 0.6990042839735244 )
( 217 , 0.6976518500178728 )
( 218 , 0.6963082155137422 )
( 219 , 0.6949734161713111 )
( 220 , 0.6936474720382179 )
( 221 , 0.6923304126385247 )
( 222 , 0.6910222688842839 )
( 223 , 0.6897230664086098 )
( 224 , 0.6884328215880011 )
( 225 , 0.68715155545897 )
( 226 , 0.685879297025733 )
( 227 , 0.6846160624112173 )
( 228 , 0.6833618788338623 )
( 229 , 0.6821167637486012 )
( 230 , 0.6808807333812892 )
( 231 , 0.6796538079522889 )
( 232 , 0.6784360048459847 )
( 233 , 0.6772273350088622 )
( 234 , 0.6760278195183982 )
( 235 , 0.6748374708289099 )
( 236 , 0.6736563059448365 )
( 237 , 0.6724843361613387 )
( 238 , 0.6713215754763809 )
( 239 , 0.6701680432246909 )
( 240 , 0.6690237370155527 )
( 241 , 0.6678886847939055 )
( 242 , 0.6667628807606308 )
( 243 , 0.6656463411133148 )
( 244 , 0.6645390731987556 )
( 245 , 0.6634410921746458 )
( 246 , 0.6623523956590445 )
( 247 , 0.661273005533208 )
( 248 , 0.6602029167322874 )
( 249 , 0.6591421352189045 )
( 250 , 0.658090680506899 )
( 251 , 0.657048545603334 )
( 252 , 0.6560157358296874 )
( 253 , 0.6549922615018781 )
( 254 , 0.6539781295256947 )
( 255 , 0.6529733285360758 )
( 256 , 0.6519778827947836 )
( 257 , 0.6509917770040934 )
( 258 , 0.6500150272339945 )
( 259 , 0.6490476168023999 )
( 260 , 0.6480895753706621 )
( 261 , 0.6471408821258181 )
( 262 , 0.6462015502348482 )
( 263 , 0.6452715745089095 )
( 264 , 0.6443509534032368 )
( 265 , 0.6434396905495547 )
( 266 , 0.6425377876576108 )
( 267 , 0.6416452385319622 )
( 268 , 0.6407620450886494 )
( 269 , 0.6398882047634912 )
( 270 , 0.6390237277257488 )
( 271 , 0.6381685954953121 )
( 272 , 0.6373228132965242 )
( 273 , 0.6364863790266116 )
( 274 , 0.635659289737813 )
( 275 , 0.6348415370882937 )
( 276 , 0.6340331380175563 )
( 277 , 0.6332340652693282 )
( 278 , 0.6324443302115342 )
( 279 , 0.6316639308534082 )
( 280 , 0.6308928495864495 )
( 281 , 0.6301311070350444 )
( 282 , 0.6293786655980216 )
( 283 , 0.6286355507299862 )
( 284 , 0.627901746504924 )
( 285 , 0.6271772532728006 )
( 286 , 0.6264620565419105 )
( 287 , 0.6257561675647361 )
( 288 , 0.6250595626740962 )
( 289 , 0.6243722502941852 )
( 290 , 0.623694222566933 )
( 291 , 0.6230254724300656 )
( 292 , 0.6223659992263787 )
( 293 , 0.6217157956962852 )
( 294 , 0.62107485938647 )
( 295 , 0.6204431658976495 )
( 296 , 0.6198207445339244 )
( 297 , 0.6192075656042605 )
( 298 , 0.6186036296835297 )
( 299 , 0.6180089265832344 )
( 300 , 0.6174234642531772 )
( 301 , 0.6168472289866276 )
( 302 , 0.6162802117720828 )
( 303 , 0.6157224085504757 )
( 304 , 0.6151738089741422 )
( 305 , 0.6146344337381904 )
( 306 , 0.6141042404425537 )
( 307 , 0.6135832564538729 )
( 308 , 0.6130714327505149 )
( 309 , 0.6125688199878632 )
( 310 , 0.6120753734739296 )
( 311 , 0.6115911085162233 )
( 312 , 0.6111159942963131 )
( 313 , 0.6106500388818474 )
( 314 , 0.6101932537014819 )
( 315 , 0.6097456164084705 )
( 316 , 0.6093071123981646 )
( 317 , 0.6088777523553991 )
( 318 , 0.6084575400843503 )
( 319 , 0.6080464456939707 )
( 320 , 0.607644477860961 )
( 321 , 0.6072516298756659 )
( 322 , 0.6068678910114662 )
( 323 , 0.6064932607808006 )
( 324 , 0.6061277402043674 )
( 325 , 0.6057713278052945 )
( 326 , 0.605423996691907 )
( 327 , 0.6050857534406137 )
( 328 , 0.6047566081501793 )
( 329 , 0.6044365414569404 )
( 330 , 0.6041255438699284 )
( 331 , 0.6038236279783992 )
( 332 , 0.6035307838933414 )
( 333 , 0.6032469882998408 )
( 334 , 0.6029722709587919 )
( 335 , 0.602706602949621 )
( 336 , 0.6024499857898091 )
( 337 , 0.6022024212847483 )
( 338 , 0.6019639120564472 )
( 339 , 0.6017344267605725 )
( 340 , 0.6015139988324095 )
( 341 , 0.6013025941176346 )
( 342 , 0.6011002290534139 )
( 343 , 0.600906891441349 )
( 344 , 0.6007225879112852 )
( 345 , 0.6005472965179767 )
( 346 , 0.6003810444002263 )
( 347 , 0.6002237988518444 )
( 348 , 0.6000755737286774 )
( 349 , 0.5999363749236754 )
( 350 , 0.5998061771745562 )
( 351 , 0.5996849902864667 )
( 352 , 0.5995728196540395 )
( 353 , 0.5994696493001649 )
( 354 , 0.5993754891053432 )
( 355 , 0.5992903297051001 )
( 356 , 0.5992141707068066 )
( 357 , 0.5991470179127677 )
( 358 , 0.5990888778905445 )
( 359 , 0.5990397180871688 )
( 360 , 0.5989995866949327 )
( 361 , 0.5989684367606773 )
( 362 , 0.5989462807139888 )
( 363 , 0.5989331257173296 )
( 364 , 0.5989289753109316 )
( 365 , 0.5989338146218454 )
( 366 , 0.5989476538427233 )
( 367 , 0.5989704924714608 )
( 368 , 0.599002331666862 )
( 369 , 0.5990431653216487 )
( 370 , 0.5990930099309947 )
( 371 , 0.5991518444655684 )
( 372 , 0.5992196768383953 )
( 373 , 0.5992965203507374 )
( 374 , 0.5993823678125569 )
( 375 , 0.5994772222627553 )
( 376 , 0.5995810674163178 )
( 377 , 0.5996939335171411 )
( 378 , 0.5998158097261189 )
( 379 , 0.5999466916173083 )
( 380 , 0.6000865897984844 )
( 381 , 0.6002355020757343 )
( 382 , 0.6003934326976297 )
( 383 , 0.600560383371675 )
( 384 , 0.600736358611098 )
( 385 , 0.6009213565489175 )
( 386 , 0.60111538313117 )
( 387 , 0.6013184406493908 )
( 388 , 0.6015305243587279 )
( 389 , 0.6017516454561799 )
( 390 , 0.601981826558691 )
( 391 , 0.6022210236984565 )
( 392 , 0.6024692754682118 )
( 393 , 0.6027265938614164 )
( 394 , 0.6029929532202272 )
( 395 , 0.6032683679282973 )
( 396 , 0.6035528453226934 )
( 397 , 0.6038463911358838 )
( 398 , 0.6041489972648796 )
( 399 , 0.6044606845027224 )
( 400 , 0.6047814505260041 )
( 401 , 0.6051112957378513 )
( 403 , 0.6057982480626899 )
( 405 , 0.6065215820729429 )
( 406 , 0.6068969062709624 )
( 407 , 0.6072813392069619 )
( 408 , 0.6076748861584448 )
( 409 , 0.6080775466869177 )
( 410 , 0.6084893391636129 )
( 411 , 0.6089102623949464 )
( 412 , 0.6093403193100975 )
( 414 , 0.6102278485748408 )
( 416 , 0.6111519854265919 )
( 417 , 0.6116278020160127 )
( 418 , 0.6121127686619742 )
( 419 , 0.6126069192734317 )
( 420 , 0.6131102454627487 )
( 421 , 0.6136227592944251 )
( 422 , 0.6141444478451483 )
( 423 , 0.6146753377946076 )
( 424 , 0.6152154338770541 )
( 425 , 0.6157647207966698 )
( 426 , 0.6163232380270798 )
( 427 , 0.6168909567377621 )
( 428 , 0.6174678994009097 )
( 429 , 0.6180540644045315 )
( 430 , 0.6186494750505944 )
( 431 , 0.6192541116658967 )
( 432 , 0.6198680003824576 )
( 433 , 0.620491130817654 )
( 434 , 0.6211235213713167 )
( 435 , 0.6217651747761452 )
( 437 , 0.6230762667682116 )
( 438 , 0.6237457265584562 )
( 439 , 0.6244244577447692 )
( 440 , 0.6251124883156848 )
( 441 , 0.6258097901520453 )
( 442 , 0.6265164068913682 )
( 444 , 0.6279575130912115 )
( 445 , 0.6286920289106698 )
( 446 , 0.6294358665272781 )
( 447 , 0.6301890031807136 )
( 448 , 0.6309514714961256 )
( 449 , 0.631723259449895 )
( 450 , 0.6325043729138634 )
( 451 , 0.6332948213938211 )
( 452 , 0.6340946013458058 )
( 454 , 0.6357221845110386 )
( 456 , 0.637387137619864 )
( 458 , 0.6390894787591225 )
( 459 , 0.6399546719565051 )
( 460 , 0.6408292355675484 )
( 461 , 0.6417131381228629 )
( 463 , 0.6435090210240597 )
( 464 , 0.6444210033362664 )
( 466 , 0.6462730319113554 )
( 467 , 0.6472130846419837 )
( 468 , 0.6481624879465498 )
( 469 , 0.6491212500987813 )
( 470 , 0.6500893663484257 )
( 471 , 0.6510668291221028 )
( 472 , 0.652053646612179 )
( 473 , 0.6530498122648563 )
( 474 , 0.6540553247288664 )
( 475 , 0.6550701696254332 )
( 477 , 0.6571278800459487 )
( 479 , 0.6592228997908222 )
( 480 , 0.6602843843321348 )
( 481 , 0.6613551844993413 )
( 482 , 0.6624352956042193 )
( 483 , 0.6635246980958944 )
( 484 , 0.6646233932592405 )
( 485 , 0.6657313609200965 )
( 486 , 0.6668486170351247 )
( 488 , 0.6691108901658329 )
( 489 , 0.6702559036000862 )
( 490 , 0.6714101418474846 )
( 491 , 0.6725736101771735 )
( 492 , 0.6737462779714026 )
( 493 , 0.6749281466076685 )
( 494 , 0.6761191900921634 )
( 495 , 0.6773194059620654 )
( 496 , 0.6785287744255877 )
( 497 , 0.6797472803636999 )
( 498 , 0.6809749012998393 )
( 499 , 0.682211625578201 )
( 500 , 0.6834574373360028 )
( 501 , 0.6847123176505249 )
( 502 , 0.6859762385055442 )
( 503 , 0.6872491880455454 )
( 504 , 0.6885311383504032 )
( 506 , 0.6911219636450491 )
( 507 , 0.6924307811977638 )
( 508 , 0.6937485262050986 )
( 509 , 0.6950751421664592 )
( 510 , 0.6964106200983977 )
( 512 , 0.6991080327168774 )
( 514 , 0.7018405347301914 )
( 516 , 0.7046078635133038 )
( 517 , 0.7060045083877708 )
( 518 , 0.7074097690411388 )
( 520 , 0.7102459462611467 )
( 521 , 0.7116767919690825 )
( 522 , 0.7131161034702405 )
( 525 , 0.717484341278889 )
( 527 , 0.7204379972580636 )
( 529 , 0.723424417549229 )
( 530 , 0.7249297952979556 )
( 532 , 0.7279646229790129 )
( 533 , 0.7294939579863099 )
( 534 , 0.73103117985274 )
( 537 , 0.7356895585040468 )
( 538 , 0.7372577111018296 )
( 540 , 0.7404167110242817 )
( 545 , 0.7484433821434016 )
( 546 , 0.750070301145296 )
( 547 , 0.7517042346218977 )
( 549 , 0.754992840553589 )
( 552 , 0.759976424199533 )
( 553 , 0.7616507710618546 )
( 559 , 0.7718289268991594 )
( 560 , 0.773546392594363 )
( 563 , 0.7787329683310542 )
( 566 , 0.7839685637365206 )
( 567 , 0.7857241407047049 )
( 568 , 0.7874846973795543 )
( 570 , 0.7910203438787998 )
( 571 , 0.7927951693617091 )
( 572 , 0.7945745140403949 )
( 574 , 0.7981462054075054 )
( 576 , 0.8017343642575991 )
( 577 , 0.8035342932011835 )
( 578 , 0.8053379288362491 )
( 580 , 0.8089557721932202 )
( 581 , 0.8107696966588043 )
( 583 , 0.8144067862209627 )
( 584 , 0.8162296650045998 )
( 585 , 0.8180552156583916 )
( 591 , 0.8290558784577777 )
( 592 , 0.8308956736145436 )
( 596 , 0.8382667362902628 )
( 600 , 0.8456481102460278 )
( 601 , 0.847493683175325 )
( 602 , 0.8493389531841155 )
( 603 , 0.8511837198011507 )
( 607 , 0.8585538941168227 )
( 608 , 0.8603931984490986 )
( 613 , 0.8695597139029089 )
( 616 , 0.8750278775252275 )
( 618 , 0.878656388419901 )
( 624 , 0.8894404351049833 )
( 625 , 0.8912205099937034 )
( 626 , 0.8929950241819322 )
( 628 , 0.896526339855149 )
( 630 , 0.9000324328291448 )
( 636 , 0.9103794435885835 )
( 640 , 0.9171144137936761 )
( 644 , 0.9236997346900447 )
( 646 , 0.9269310781432699 )
( 648 , 0.9301187118585125 )
( 653 , 0.9378829364582941 )
( 654 , 0.9393985608654517 )
( 655 , 0.9409011426469833 )
( 657 , 0.9438662248279978 )
( 659 , 0.946776078684087 )
( 661 , 0.9496286392848992 )
( 665 , 0.9551536196369096 )
( 666 , 0.956495854736731 )
( 676 , 0.9689916212711063 )
( 679 , 0.9723899906993297 )
( 689 , 0.9824375893087425 )
( 690 , 0.9833286681614295 )
( 692 , 0.9850463156314929 )
( 693 , 0.9858724972600061 )
( 700 , 0.9910308720624721 )
( 701 , 0.9916766021966207 )
( 703 , 0.9928982117925256 )
( 704 , 0.9934738163064846 )
( 715 , 0.9982216511345522 )
( 716 , 0.9985067847798336 )
( 725 , 0.9999535914381598 )
( 726 , 0.9999892898610176 )
( 728 , 0.9999854552831714 )
( 730 , 0.9998813177240827 )
( 732 , 0.9996769415688491 )
( 739 , 0.9981758707647151 )
( 741 , 0.9975240830585531 )
( 746 , 0.9954668625866058 )
( 757 , 0.9888444458275278 )
( 758 , 0.9881037644201439 )
( 771 , 0.976494309104758 )
( 772 , 0.975455463435481 )
( 778 , 0.9688139167381212 )
( 783 , 0.9627690124271598 )
( 784 , 0.9615071926909527 )
( 788 , 0.9562916775642111 )
( 791 , 0.9522104443426589 )
( 814 , 0.9168596243362866 )
( 819 , 0.9084128318315523 )
( 831 , 0.8873816001295584 )
( 834 , 0.8819936304413385 )
( 836 , 0.8783795358699199 )
( 844 , 0.8637859085764107 )
( 845 , 0.8619499331171311 )
( 847 , 0.8582725496901183 )
( 855 , 0.8435202656193758 )
( 861 , 0.8324552604156142 )
( 863 , 0.8287747247934408 )
( 864 , 0.826936710031321 )
( 866 , 0.8232660435510855 )
( 877 , 0.8032588524819854 )
( 885 , 0.7889799119627791 )
( 886 , 0.7872152079272439 )
( 900 , 0.7630741826688989 )
( 901 , 0.7613943616260536 )
( 907 , 0.751453965914333 )
( 913 , 0.7417637391864993 )
( 921 , 0.7292596259077949 )
( 925 , 0.7231949680767242 )
( 926 , 0.7216989306591838 )
( 932 , 0.7128955037876514 )
( 936 , 0.7071943564173598 )
( 937 , 0.7057904183991992 )
( 942 , 0.6989005846791049 )
( 944 , 0.6962058656664344 )
( 945 , 0.6948717376773034 )
( 952 , 0.6857824083180122 )
( 968 , 0.666677192970746 )
( 981 , 0.6528969015965435 )
( 988 , 0.6461301301768038 )
( 990 , 0.6442809609297895 )};

\addplot[blue, mark=none] coordinates{(1.0,0.9999999937348335)
(2.0,0.9999992100430583)
(3.0,0.9999968723686782)
(4.0,0.9999929559222869)
(5.0,0.9999874911030379)
(6.0,0.9999804485680988)
(7.0,0.9999718600534672)
(8.0,0.9999617026704507)
(9.0,0.9999499834020773)
(10.0,0.9999367026301968)
(11.0,0.9999218763295191)
(12.0,0.9999054852034915)
(13.0,0.999887550020131)
(14.0,0.9998680565859511)
(15.0,0.9998470142649165)
(16.0,0.9998244148837307)
(17.0,0.9998002873294771)
(18.0,0.9997746068333451)
(19.0,0.9997473965833406)
(20.0,0.9997186360050594)
(21.0,0.9996883459137701)
(22.0,0.9996565332569903)
(23.0,0.9996231888744517)
(24.0,0.9995883242039808)
(25.0,0.9995519405581955)
(26.0,0.9995140316093509)
(27.0,0.9994746124367482)
(28.0,0.999433700229586)
(29.0,0.9993912862485684)
(30.0,0.9993473630865843)
(31.0,0.9993019394879417)
(32.0,0.9992550326909478)
(33.0,0.9992066465478604)
(34.0,0.9991567815636158)
(35.0,0.9991054508718199)
(36.0,0.9990526396698836)
(37.0,0.9989983731170363)
(38.0,0.9989426518419614)
(39.0,0.998885477407186)
(40.0,0.9988268515702522)
(41.0,0.9987667964640659)
(42.0,0.9987053135728922)
(43.0,0.9986424017704627)
(44.0,0.9985780595385805)
(45.0,0.9985123166597222)
(46.0,0.9984451781003985)
(47.0,0.9983766346053923)
(48.0,0.9983067051584651)
(49.0,0.9982353779042303)
(50.0,0.9981627048444965)
(51.0,0.9980886459940829)
(52.0,0.9980132316149903)
(53.0,0.9979364687686197)
(54.0,0.9978583518935967)
(55.0,0.9977789180924983)
(56.0,0.9976981513389246)
(57.0,0.99761604942474)
(58.0,0.9975326628106024)
(59.0,0.9974479688486624)
(60.0,0.9973619938105087)
(61.0,0.9972747236053336)
(62.0,0.9971861824577108)
(63.0,0.9970963861198686)
(64.0,0.9970053194253691)
(65.0,0.996913026868023)
(66.0,0.9968194885327467)
(67.0,0.9967247384159367)
(69.0,0.9965315874384691)
(70.0,0.9964332374853534)
(71.0,0.9963336842445316)
(72.0,0.9962329524101488)
(74.0,0.9960280421680181)
(75.0,0.9959238582958323)
(76.0,0.9958185589376614)
(77.0,0.9957121162880024)
(78.0,0.9956046023688538)
(79.0,0.9954959687522045)
(81.0,0.9952754739494538)
(82.0,0.9951636278901503)
(84.0,0.9949367999454449)
(85.0,0.9948218339981627)
(86.0,0.9947058685459284)
(88.0,0.9944709219534618)
(89.0,0.9943519835743849)
(90.0,0.9942320754947689)
(91.0,0.9941112091764254)
(92.0,0.9939894178883215)
(93.0,0.993866698464101)
(94.0,0.9937430579208213)
(95.0,0.9936185197861264)
(96.0,0.9934930899506279)
(97.0,0.9933667899018418)
(99.0,0.9931115983827274)
(101.0,0.9928530699814141)
(102.0,0.9927225931146227)
(103.0,0.9925913230945665)
(104.0,0.992459255290096)
(106.0,0.9921928452764299)
(108.0,0.9919234684597086)
(109.0,0.9917877066936049)
(110.0,0.991651254628376)
(111.0,0.9915141012303734)
(112.0,0.9913762846982607)
(113.0,0.9912378066481853)
(114.0,0.9910986902517926)
(115.0,0.990958938838189)
(116.0,0.9908185791179398)
(117.0,0.990677631577571)
(118.0,0.9905360746929066)
(119.0,0.9903939554522558)
(120.0,0.9902512884548134)
(121.0,0.9901080736029285)
(122.0,0.9899643333953176)
(123.0,0.9898200736317392)
(124.0,0.9896753201561367)
(125.0,0.9895300806026481)
(127.0,0.9892382310014416)
(128.0,0.9890916232386173)
(129.0,0.9889446067120352)
(130.0,0.9887971757273049)
(131.0,0.9886493486985992)
(133.0,0.9883525641827555)
(134.0,0.98820364631453)
(135.0,0.9880543920768428)
(136.0,0.9879048204319382)
(137.0,0.9877549348526318)
(138.0,0.9876047594107341)
(139.0,0.9874543362502949)
(140.0,0.9873036336861732)
(141.0,0.9871526834828905)
(142.0,0.9870015045237495)
(144.0,0.9866985262863462)
(145.0,0.9865467425058089)
(146.0,0.9863948154096335)
(147.0,0.9862427036728172)
(148.0,0.9860904829987541)
(150.0,0.9857856548894085)
(151.0,0.9856330887591381)
(152.0,0.9854804650555463)
(153.0,0.9853277621411081)
(154.0,0.9851750031677006)
(155.0,0.9850222246307806)
(156.0,0.9848694350695768)
(157.0,0.9847166271681558)
(159.0,0.9844110707664835)
(160.0,0.9842583530686416)
(161.0,0.9841057022114508)
(162.0,0.9839531108814584)
(163.0,0.9838006166379721)
(164.0,0.9836482125665628)
(165.0,0.9834959341224687)
(166.0,0.9833437844484337)
(167.0,0.9831917866536828)
(168.0,0.9830399404653754)
(169.0,0.982888279953845)
(170.0,0.982736812544735)
(171.0,0.9825855411193027)
(172.0,0.9824344929998314)
(173.0,0.9822836865878393)
(174.0,0.982133141288135)
(175.0,0.9819828318777587)
(176.0,0.9818328290127529)
(177.0,0.9816831045328829)
(178.0,0.9815336968881119)
(179.0,0.9813846133244312)
(180.0,0.9812358541100521)
(181.0,0.9810874543562892)
(182.0,0.9809394316310973)
(183.0,0.9807917840769032)
(184.0,0.9806445193897253)
(185.0,0.9804976715916827)
(186.0,0.9803512460488809)
(187.0,0.9802052510661342)
(188.0,0.9800597197527117)
(192.0,0.9794823428585033)
(193.0,0.9793392754998655)
(194.0,0.9791967254157004)
(196.0,0.9789132971352956)
(197.0,0.9787724300124795)
(198.0,0.9786321516265695)
(199.0,0.9784924812852555)
(201.0,0.9782149859502514)
(204.0,0.9778035939411068)
(206.0,0.9775326941462763)
(208.0,0.9772646248146903)
(209.0,0.9771316599127565)
(210.0,0.9769994631530623)
(211.0,0.9768680082816684)
(212.0,0.9767373251579504)
(213.0,0.9766074253625064)
(214.0,0.9764783045905653)
(216.0,0.9762225019455824)
(218.0,0.9759700128183889)
(219.0,0.9758450401505393)
(221.0,0.9755976799798575)
(223.0,0.9753538753736958)
(224.0,0.9752333191989262)
(226.0,0.9749950007402455)
(227.0,0.9748772515314096)
(228.0,0.9747604482603319)
(229.0,0.9746446174264459)
(230.0,0.9745297467630594)
(231.0,0.9744158723201943)
(232.0,0.9743029895982838)
(233.0,0.9741911163997712)
(235.0,0.9739704197238137)
(236.0,0.9738616226027202)
(237.0,0.9737538722734592)
(238.0,0.9736471797968347)
(239.0,0.9735415486830076)
(241.0,0.9733335315776143)
(242.0,0.9732311618849963)
(243.0,0.9731298977364267)
(244.0,0.9730297512285924)
(246.0,0.9728328233486326)
(247.0,0.972736063158887)
(250.0,0.9724527285204457)
(251.0,0.9723606274524038)
(255.0,0.9720041277579706)
(256.0,0.9719180299606541)
(258.0,0.971749507166237)
(260.0,0.9715859398267873)
(261.0,0.9715060436119662)
(263.0,0.9713500216522469)
(265.0,0.9711991076636172)
(266.0,0.9711255876958992)
(268.0,0.9709824414996165)
(270.0,0.97084452704697)
(273.0,0.9706475795945616)
(274.0,0.9705846035399107)
(275.0,0.9705229681535851)
(276.0,0.9704626882718829)
(279.0,0.9702899977291809)
(280.0,0.9702351783727873)
(282.0,0.970129657126631)
(283.0,0.9700789818954288)
(285.0,0.9699817872134195)
(286.0,0.9699353008379711)
(287.0,0.9698902084522006)
(288.0,0.9698465188357762)
(290.0,0.9697633891931096)
(291.0,0.969723954818819)
(292.0,0.9696859390058851)
(293.0,0.9696493415203558)
(295.0,0.9695804594756876)
(297.0,0.9695173053333437)
(298.0,0.9694878974355083)
(299.0,0.9694599267367265)
(300.0,0.9694333940199032)
(301.0,0.9694083264266639)
(302.0,0.9693846926521534)
(303.0,0.9693625256856069)
(304.0,0.9693418106292977)
(305.0,0.9693225472274869)
(307.0,0.9692884083485315)
(308.0,0.9692735267714327)
(309.0,0.9692601155523953)
(310.0,0.9692481602090068)
(312.0,0.9692286471796766)
(313.0,0.9692210917422895)
(314.0,0.9692150092889636)
(315.0,0.9692103914240279)
(316.0,0.9692072347315801)
(318.0,0.9692053407015725)
(319.0,0.969206593505365)
(320.0,0.9692093217897944)
(321.0,0.9692135217676758)
(322.0,0.9692191754016911)
(323.0,0.9692263137205818)
(324.0,0.969234902362941)
(325.0,0.9692449677053556)
(326.0,0.9692565013920889)
(328.0,0.9692839394642546)
(329.0,0.9692998683683536)
(331.0,0.9693360836329721)
(333.0,0.969378127366534)
(334.0,0.969401333741687)
(336.0,0.9694520864327087)
(338.0,0.969508634792444)
(341.0,0.9696042582620557)
(342.0,0.9696390107445856)
(344.0,0.9697127923193813)
(346.0,0.9697922585515395)
(347.0,0.969834125391758)
(348.0,0.9698774088062622)
(349.0,0.9699220932237287)
(351.0,0.9700156665223495)
(353.0,0.9701148326416122)
(354.0,0.9701665036877609)
(359.0,0.9704454779764682)
(362.0,0.9706291913317776)
(363.0,0.9706931239987071)
(364.0,0.9707583916121094)
(366.0,0.9708928922799335)
(367.0,0.9709621313239913)
(368.0,0.9710326790748884)
(369.0,0.9711045366182957)
(370.0,0.9711776766192657)
(371.0,0.9712521244320745)
(372.0,0.9713278496112535)
(373.0,0.9714048495232507)
(374.0,0.9714831275832972)
(375.0,0.9715626684593507)
(376.0,0.9716434638191656)
(377.0,0.9717255059376843)
(380.0,0.9719790550576367)
(381.0,0.9720660335047542)
(383.0,0.9722435881514929)
(386.0,0.9725188636971314)
(388.0,0.9727082500117475)
(389.0,0.9728046767274459)
(390.0,0.9729022441309757)
(392.0,0.9731007644687749)
(395.0,0.9734069166445328)
(396.0,0.9735111534555243)
(400.0,0.973938794886764)
(401.0,0.9740483281211227)
(402.0,0.9741588866268408)
(403.0,0.9742704788669793)
(405.0,0.974496653896831)
(407.0,0.9747267915987564)
(408.0,0.9748433142895866)
(414.0,0.975562147271018)
(415.0,0.9756851309327739)
(419.0,0.9761857279641393)
(420.0,0.9763129782625818)
(421.0,0.976441051270357)
(422.0,0.9765699297985642)
(423.0,0.9766996118906939)
(424.0,0.976830072528349)
(425.0,0.9769613002000976)
(427.0,0.9772260271703294)
(429.0,0.9774936787135039)
(430.0,0.9776285820314539)
(431.0,0.9777641849862455)
(432.0,0.9779004550346546)
(434.0,0.9781750094119542)
(435.0,0.9783132507202497)
(437.0,0.9785916315021836)
(438.0,0.9787317256538236)
(440.0,0.9790136974442055)
(441.0,0.9791555330219824)
(445.0,0.9797282471247934)
(447.0,0.980017635998435)
(448.0,0.980163048737518)
(449.0,0.9803088983779592)
(450.0,0.9804552024202672)
(453.0,0.9808966098256607)
(455.0,0.9811928312811017)
(457.0,0.9814904763637081)
(458.0,0.9816397782431685)
(459.0,0.9817894168417599)
(460.0,0.9819393456745612)
(461.0,0.9820895685483773)
(462.0,0.9822400337066038)
(463.0,0.9823907868248566)
(464.0,0.9825417466523205)
(465.0,0.9826929720146187)
(468.0,0.9831477798494992)
(469.0,0.9832997320828356)
(470.0,0.9834518400147105)
(471.0,0.9836040877535283)
(472.0,0.9837564573533044)
(474.0,0.9840614903995883)
(475.0,0.9842141294583562)
(476.0,0.9843668337079935)
(477.0,0.9845195893992925)
(478.0,0.9846723576152853)
(482.0,0.98528350879098)
(484.0,0.9855888834889629)
(486.0,0.9858939641698552)
(493.0,0.9869576685948314)
(494.0,0.987108909425233)
(495.0,0.9872599154016235)
(497.0,0.9875612226994375)
(498.0,0.9877114699940639)
(501.0,0.9881604368199475)
(503.0,0.9884581323932204)
(507.0,0.9890490829822642)
(508.0,0.9891958012087632)
(511.0,0.9896332856268496)
(512.0,0.9897781917497839)
(513.0,0.9899226012199266)
(514.0,0.9900664796692125)
(515.0,0.9902098483651188)
(516.0,0.9903526855225374)
(517.0,0.9904949594302204)
(522.0,0.9911975671750357)
(523.0,0.9913362396608165)
(524.0,0.9914742411358025)
(525.0,0.9916115969186726)
(529.0,0.9921540115247903)
(530.0,0.9922878123243019)
(534.0,0.9928153626492984)
(535.0,0.992945270493714)
(540.0,0.993582266226792)
(541.0,0.9937070683230159)
(542.0,0.9938309684190615)
(544.0,0.9940760216911314)
(548.0,0.9945548097753838)
(550.0,0.9947883393667863)
(552.0,0.9950178322499508)
(553.0,0.9951310228408268)
(555.0,0.9953542719448008)
(556.0,0.995464296308688)
(557.0,0.9955732389803107)
(559.0,0.9957878418667179)
(560.0,0.9958934705134607)
(562.0,0.9961013313719979)
(563.0,0.9962035607722474)
(565.0,0.9964045152856555)
(566.0,0.996503221991243)
(568.0,0.9966970666366177)
(570.0,0.9968860613101219)
(572.0,0.9970701274408028)
(576.0,0.9974231991209436)
(578.0,0.9975920321077308)
(579.0,0.9976744993934655)
(581.0,0.9978354801159921)
(582.0,0.9979139769040037)
(588.0,0.9983565160433888)
(589.0,0.9984254708953539)
(591.0,0.998559158553894)
(592.0,0.9986239001917467)
(596.0,0.9988686438610104)
(597.0,0.9989262346451887)
(598.0,0.9989823800729754)
(599.0,0.9990370723719597)
(601.0,0.9991420712600078)
(602.0,0.9991923557963576)
(603.0,0.9992411808956501)
(606.0,0.9993787082207648)
(607.0,0.999421566890217)
(612.0,0.9996132478820134)
(614.0,0.9996792943498629)
(616.0,0.9997392190569045)
(617.0,0.9997668740307736)
(619.0,0.9998175929509131)
(620.0,0.9998406274335699)
(621.0,0.9998621179158902)
(623.0,0.9999004493827853)
(624.0,0.9999172915941122)
(627.0,0.999958469102074)
(632.0,0.999995900764617)
(636.0,0.9999977128885531)
(637.0,0.9999942567093033)
(638.0,0.9999892363443379)
(640.0,0.9999745108873832)
(642.0,0.9999535409651366)
(648.0,0.9998532655019925)
(651.0,0.999782195351709)
(656.0,0.999633006323673)
(658.0,0.9995626341787315)
(659.0,0.9995251795099301)
(660.0,0.9994861926592006)
(662.0,0.999403719330471)
(664.0,0.9993152591905795)
(665.0,0.9992687809781351)
(666.0,0.9992208220071042)
(670.0,0.9990142371246029)
(671.0,0.9989589314447362)
(672.0,0.9989021824081562)
(674.0,0.9987843477935954)
(676.0,0.9986607683790057)
(677.0,0.9985968525032383)
(679.0,0.9984647677021159)
(680.0,0.9983966413834798)
(681.0,0.998327108923588)
(682.0,0.9982561848404679)
(691.0,0.9975569587605777)
(692.0,0.99747264736495)
(694.0,0.9973001338536247)
(697.0,0.9970318288753409)
(698.0,0.996939899236411)
(703.0,0.9964618554226087)
(705.0,0.9962622613225594)
(708.0,0.995954157432365)
(709.0,0.9958491813636972)
(710.0,0.9957430773920876)
(711.0,0.9956358605893979)
(712.0,0.9955275524825727)
(714.0,0.9953076608327809)
(715.0,0.9951961417374515)
(716.0,0.9950835411776844)
(717.0,0.9949699091998467)
(722.0,0.994386540014489)
(727.0,0.9937789679182948)
(732.0,0.9931487788096214)
(734.0,0.9928907297700496)
(735.0,0.9927604700197695)
(739.0,0.9922316285070534)
(740.0,0.9920975136986244)
(741.0,0.9919626778764948)
(745.0,0.9914162827991805)
(746.0,0.9912780017217347)
(747.0,0.9911390621089086)
(750.0,0.9907185187838276)
(753.0,0.9902926836151614)
(755.0,0.9900060282261535)
(762.0,0.9889872386618555)
(764.0,0.9886922070850703)
(766.0,0.9883956433845192)
(767.0,0.9882468167353146)
(771.0,0.9876482990812271)
(772.0,0.9874979469619939)
(773.0,0.9873473102513591)
(774.0,0.9871964330173033)
(775.0,0.9870453144998615)
(782.0,0.9859822664761451)
(784.0,0.985677307879031)
(786.0,0.9853720081775234)
(788.0,0.985066499645713)
(789.0,0.9849136830168425)
(790.0,0.9847608944457356)
(795.0,0.9839973137827592)
(799.0,0.9833878469437691)
(800.0,0.9832358085963078)
(802.0,0.9829321977091008)
(809.0,0.9818762574643676)
(810.0,0.9817264488905493)
(814.0,0.9811304181307152)
(819.0,0.9803936155984879)
(823.0,0.9798118776452172)
(826.0,0.9793806613033282)
(829.0,0.9789542067479721)
(831.0,0.9786727324799389)
(833.0,0.9783936483815596)
(835.0,0.9781170469978768)
(841.0,0.9773032665774107)
(844.0,0.9769060153092923)
(846.0,0.9766449740348743)
(847.0,0.9765156285188527)
(848.0,0.9763870814925457)
(850.0,0.9761324494385771)
(851.0,0.9760063744001849)
(860.0,0.974911265928305)
(864.0,0.974448756193831)
(866.0,0.9742234196790831)
(868.0,0.9740021351743328)
(876.0,0.9731591124963469)
(877.0,0.973058637805707)
(879.0,0.9728610549487074)
(880.0,0.9727639784037486)
(881.0,0.9726680344115198)
(882.0,0.9725732517808924)
(883.0,0.9724796368851221)
(886.0,0.9722058506755795)
(891.0,0.9717736129948601)
(895.0,0.9714500455044398)
(896.0,0.9713723075101451)
(900.0,0.9710741576945607)
(901.0,0.9710028563125591)
(904.0,0.9707968171244703)
(905.0,0.9707307813327686)
(908.0,0.9705406892033797)
(909.0,0.9704800118238512)
(911.0,0.9703627359528467)
(912.0,0.9703061451255579)
(913.0,0.9702509197048458)
(919.0,0.9699486261787169)
(924.0,0.9697352345850808)
(927.0,0.96962422200341)
(928.0,0.9695900831860346)
(930.0,0.9695260981096686)
(934.0,0.9694154331536428)
(939.0,0.9693097615991006)
(943.0,0.9692514760036652)
(945.0,0.9692311100228941)
(946.0,0.9692231345261968)
(947.0,0.9692166252873808)
(950.0,0.9692058961916465)
(952.0,0.9692060791027267)
(955.0,0.9692173863414235)
(958.0,0.9692419059261178)
(960.0,0.9692655744638273)
(962.0,0.9692951066471236)
(965.0,0.9693503419074143)
(972.0,0.969530106463531)
(974.0,0.9695944697075838)
(975.0,0.9696287982972904)
(981.0,0.9698647203948848)
(982.0,0.9699090017493315)
(986.0,0.9701001287025447)
(987.0,0.9701513948351369)
(988.0,0.970204043678791)
(994.0,0.9705487297900047)
(997.0,0.9707393444597182)};
\end{axis}
\end{tikzpicture}
\caption{Plot showing, for $\omega = 0.01$, the concurrence of the bipartite system $ q q'$ as a function of time, illustrating the entanglement dynamics. Notably, the concurrence exhibits slight oscillations at $\omega= 8 \Gamma $ 
(blue curve), and more pronounced oscillations at \(\omega=2\Gamma\) (black curve), indicative of entanglement non-monotonicity near the exceptional point in the $PT$-symmetric regime. By contrast, for \(\omega=0.89\Gamma\) (red curve), close to the EP and in the $PT$ broken phase, the concurrence of the bipartite system decays, displaying a more traditional open-system entanglement dynamics.}
\label{fig:concurrence_exp_results}
\end{figure}
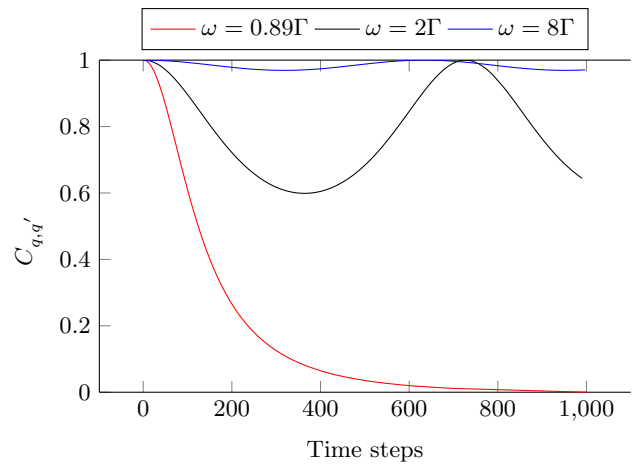

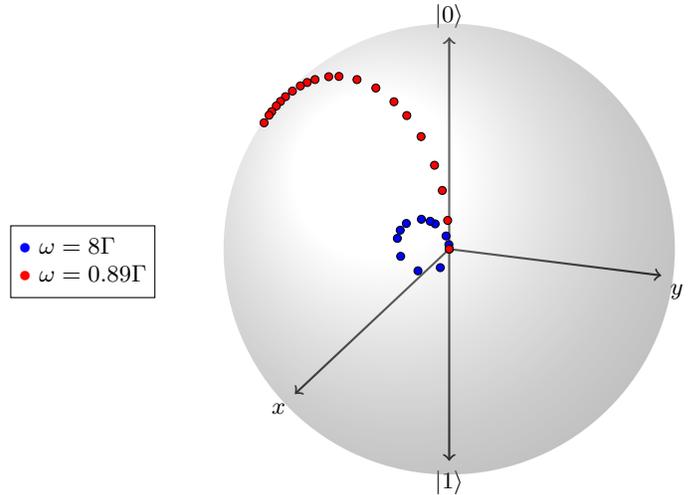
\begin{figure}
    \tdplotsetmaincoords{70}{110} 
    \begin{tikzpicture}[tdplot_main_coords, scale=3] 
        \draw[thick,->] (0,0,0) -- (2,0,0) node[anchor=north east]{$x$}; 
        \draw[thick,->] (0,0,0) -- (0,1,0) node[anchor=north west]{$y$}; 
        \draw[thick,->] (0,0,0) -- (0,0,1) node[anchor=south]{$|0\rangle$}; 
        \draw[thick,->] (0,0,0) -- (0,0,-1) node[anchor=north]{$|1\rangle$}; 

        \shade[ball color = white!, opacity = 0.4] (0,0,0) circle (1cm); 

        \foreach \point in {
            (0.0, 0.0, 2.220446049250313e-16), (3.587804868113044e-15, -0.014984782636294686, 0.059954173166140046), (5.324216993155856e-15, -0.06620938674889959, 0.11065305031357908), (1.1686445711456284e-14, -0.13135645064238632, 0.12442355012154294), (1.3036984924656583e-14, -0.2022819374631445, 0.09565501744832278), (2.2173613726628515e-14, -0.2447217968368134, 0.020113123414667033), (2.9573814613447095e-14, -0.22898158623024953, -0.06276970735580045), (3.7204469142016046e-14, -0.14742679782310897, -0.12103158409747683), (4.294785873983183e-14, -0.0421404265086498, -0.09284571929879193), (5.670483102400661e-14, -0.0016375688554553375, 0.02080498619040333), (5.661658739322359e-14, -0.08982491594055891, 0.12007990185256262), (6.341018382704584e-14, -0.23101019135914397, 0.06040917668024026)
        }{

    \draw[fill=blue] \point circle (0.5pt);
}
    \foreach \point in {
        (0.0, 0.0, 2.220446049250313e-16),
(1.2462376530742222e-15, -0.00738741855221299, 0.1344039259013381),
(1.1222430996138566e-15, -0.03262210643351238, 0.27242365584870637),
(2.935558204435755e-15, -0.06936401615851548, 0.38695070635743894),
(2.9216656754192125e-15, -0.13267918860215558, 0.5143218058208444),
(1.9790859353973488e-15, -0.2002676386202865, 0.6052449055790953),
(3.525320673729908e-15, -0.2606543021675891, 0.6628474255260453),
(5.2326186417140075e-15, -0.3460399863853036, 0.7169339878073147),
(6.8966504369156845e-15, -0.43499686459889736, 0.7457720036257127),
(7.595600918733106e-15, -0.5199804088347146, 0.7501158491252826),
(9.20357030299875e-15, -0.5681887598096667, 0.7427703565009225),
(1.2955901819399366e-14, -0.6329237198136961, 0.7213359236410513),
(1.1615904193811106e-14, -0.6698599179487614, 0.7027618850274017),
(1.464262822568853e-14, -0.7019560653566684, 0.6825342231089444),
(1.7392538840032557e-14, -0.739488328588571, 0.6535681585766884),
(1.7279158827020534e-14, -0.7720976778763984, 0.6231670968839274),
(1.9467115669426973e-14, -0.7957338717126226, 0.5976545165059616),
(2.08629352769098e-14, -0.8154680474845618, 0.5738020242024038),
(2.250412991941505e-14, -0.8372344961919365, 0.5444134071100536),
(2.3604360881216972e-14, -0.8497340033099678, 0.5258768771742126),
(2.717263195619654e-14, -0.8731786130020154, 0.48727084028063095)    }{
        \draw[fill=red] \point circle (0.5pt);
    }
\node[draw, fill=white] at (2, -1, 0.5) { 
    \begin{tabular}{@{}r@{ }l@{}}

        \textcolor{blue}{$\bullet$} & $\omega = 8\Gamma$ \\
        \textcolor{red}{$\bullet$} & $\omega= 0.89\Gamma$ 

    \end{tabular}
};
\end{tikzpicture}
\caption{The Bloch sphere representation of the system qubit's reduced density matrix as it evolves under $H_{\text{eff}}$, for the same protocol followed in Fig.~\ref{fig:concurrence_exp_results}.
At time $t=0$ the pair $q q'$ is initialized in a Bell state; hence the reduced density matrix for qubit $q$ is maximally mixed. The red points are in the $PT$ broken phase, where the entanglement decays (see Fig.~\ref{fig:concurrence_exp_results}). This results in the red points leaving the origin and evolving to a pure state at the surface of the sphere, leading to state purification. The blue points are in the $PT$ symmetric phase, where the entanglement is strong but exhibits small oscillations.}
\label{fig:blochsphere}
\end{figure}

The simulation uses three qubits, two for the Bell pair $ q q'$ and a third for an ancilla to simulate $H_{\rm eff}$ as described in Fig.~\ref{fig:model}.
The results (from classical simulations) are shown in Fig.~\ref{fig:concurrence_exp_results}. 
For a  complementary view, 
Fig.~\ref{fig:blochsphere} shows the Bloch sphere coordinates for the non-Hermitian qubit's reduced density matrix, showing the purification 
in the $PT$-broken phase (red points).
The observed entanglement oscillation in the $PT$ symmetric phase violates  \cite{PhysRevA.90.054301, PhysRevLett.119.190401}  the entanglement monotonicity principle holding that entanglement cannot be increased under the action of local (single-qubit) completely-positive trace preserving (CPTP) maps. However, density matrix time evolution
\begin{eqnarray}
\rho \mapsto  \frac{ e^{-i H_{\rm eff} t } 
\rho e^{i H^\dagger_{\rm eff} t } }{{\rm tr} (e^{-i H_{\rm eff} t }  \rho e^{i H^\dagger_{\rm eff} t }) },
\end{eqnarray}
falls outside the space of linear CPTP maps
when $H_{\rm eff}  \neq H_{\rm eff} ^\dagger$
(due to the required normalization),
enabling this striking phenomena.
 
\section{\label{sec:concl}Conclusion}

We have used a nonunitary (postselected) variational quantum circuit to simulate a qubit evolving according to the non-Hermitian Hamiltonian (\ref{Heff def}). 
The effects of $H_{\rm eff}$ on single-qubit and two-qubit dynamics have been explored.
Our work suggests interesting avenues for further investigation:
(i) How to simulate multi-qubit non-Hermitian Hamiltonians? One approach has been  
proposed by Zhang {\it et al.} \cite{Zhang2021}. 
(ii) More broadly, can non-Hermitian effects be used to enhance quantum information processing by expanding the allowed operations beyond linear CPTP channels?

\section{\label{sec:ackno}Acknowledgments}

This work was partly supported by the NSF under grant no. DGE-2152159. It was also supported with resources and technical expertise from the Georgia Advanced Computing Resource Center, a partnership between the University of Georgia’s Office of the Vice President for Research and the Office of the Vice President for Information Technology. We acknowledge use of the IBM Quantum Experience for this work. The views expressed are those of the authors and do not
reflect the official policy or position of IBM or the IBM
Quantum Experience team.

\FloatBarrier
\bibliography{paper}
\appendix

\end{document}